\documentclass[12pt]{article}
\begin{document}
\begin{titlepage}
\begin{flushright}
hep-th/0503224\\
TIT/HEP-536\\
TU-740\\
March, 2005\\
\end{flushright}
\vspace{0.5cm}
\begin{center}
{\Large \bf Non(anti)commutative ${\cal N}=(1,1/2)$ \\
Supersymmetric $U(1)$ Gauge Theory
}
\lineskip .75em
\vskip2.5cm
{\large Takeo Araki}${}^{1}$,\ \
{\large Katsushi Ito}${}^{2}$ \
and \ {\large Akihisa Ohtsuka}${}^{2}$
\vskip 2.5em
${}^{1}$ {\large\it Department of Physics\\
Tohoku University\\
Sendai, 980-8578, Japan}  \vskip 1.5em
${}^{2}${\large\it Department of Physics\\
Tokyo Institute of Technology\\
Tokyo, 152-8551, Japan}  \vskip 4.5em
\end{center}
\begin{abstract}
We study a reduction of deformation parameters 
in non(anti)commutative ${\cal N}=2$ harmonic superspace to 
those in non(anti)commutative ${\cal N}=1$ superspace.
By this reduction we obtain the exact gauge and supersymmetry
 transformations in the Wess-Zumino gauge of non(anti)commutative 
${\cal N}=2$ supersymmetric $U(1)$ gauge theory 
defined in the deformed harmonic superspace. 
We also find that 
the action with the first order correction in the deformation parameter 
reduces to the one in the ${\cal N}=1$ superspace by some field
redefinition. 
We construct deformed ${\cal N}=(1,1/2)$ supersymmetry in ${\cal N}=2$
supersymmetric $U(1)$  gauge theory in non(anti)commutative ${\cal N}=1$
superspace.

\end{abstract}
\end{titlepage}

\baselineskip=0.7cm
\section{Introduction}
Supersymmetric field theories in deformed
superspace\cite{ncsuper,KlPeTa} 
have been 
recently attracted much attentions in view of  studying superstring
effective field theories on the D-branes with graviphoton background
\cite{OoVa,BeSe,DeGrNi}.
Non(anti)commutative superspace is a deformed superspace with 
nonanticommutative Grassmann coordinates and the $*$-product.
Field theories in non(anti)commutative superspace is 
constructed in terms of superfields
and have been extensively studied 
\cite{Se}\cite{ArItOh1}\cite{Pert}--\cite{FeIvLeSoZu}.

${\cal N}=1$ super Yang-Mills theory 
in the non(anti)commutative ${\cal N}=1$ superspace 
has been formulated in \cite{Se}, 
where the deformation preserves the chiral structure of the theory. 
As in the commutative case, one can take the Wess-Zumino(WZ) gauge for
vector superfields to write down the deformed action in terms of  component 
fields.
Since supersymmetry transformation cannot keep the WZ gauge, one may
perform additional gauge transformation to recover the WZ gauge.
In the nonanticommutative case, this gauge transformation induces the 
terms which depends on the deformation parameter $C$.
Therefore, even if we redefine the component fields such that these  fields 
transform canonically under the gauge transformation, the supersymmetry
transformations receive the deformation due to non(anti)commutativity.

In a previous paper \cite{ArItOh1}, we wrote down the deformed action of
${\cal N}=1$ supersymmetric $U(N)$ gauge theory
with matter fields in both fundamental and adjoint representations.
We also examined invariance of the
action under the ${\cal N}=1/2$ supersymmetry transformation.
In particular, in the case of adjoint matter fields,  
we claimed that we could not find 
the (deformed) extended supersymmetry transformation 
and concluded that only the ${\cal N}=1/2$ supersymmetry is preserved. 
It was pointed out in \cite{IvLeZu}, however, that 
this conclusion had discrepancy with the general argument 
concerning the symmetry of the deformed ${\cal N}=2$ superspace whose 
Poisson structure constructed by the chiral supercharges $Q_\alpha^i$ 
preserves at least the ${\cal N}=(1,0)$ supersymmetry of the theory 
generated by $Q^1_\alpha$, $Q^2_\alpha$. 
Moreover, it is natural to expect 
that the action defined in the deformed harmonic superspace  leads to
the one in the deformed ${\cal N}=1$ superspace by 
the reduction of deformation parameters of the harmonic superspace to
the deformed ${\cal N}=1$ superspace. 
In this case, it was also argued in \cite{IvLeZu} 
that the deformation preserves the ${\cal N}=(1,1/2)$ supersymmetry 
generated by $Q^1$, $Q^2$ and $\bar{Q}_2$.
The existence of ${\cal N}=(1,1/2)$ supersymmetry is partly supported by
our recent work\cite{ArItOh3}, in which the deformed ${\cal N}=(1,0)$ 
supersymmetry has been constructed in the ${\cal N}=2$ supersymmetric
$U(1)$ gauge theory in non(anti)commutative harmonic superspace with
generic (non-singlet) deformations. 
Thus we expect ${\cal N}=(1,1/2)$ supersymmetry in the 
${\cal N}=2$ supersymmetric $U(N)$ gauge theory in the
non(anti)commutative ${\cal N}=1$ superspace.

In this paper, we will construct deformed ${\cal N}=(1,1/2)$ 
supersymmetry in ${\cal N}=2$
supersymmetric $U(1)$  gauge theory in non(anti)commutative ${\cal N}=1$
superspace.
We compare this theory to  
the one in non(anti) commutative ${\cal N}=2$ harmonic superspace obtained 
by the reduction to the deformed ${\cal N}=1$ superspace.
By this reduction we find  the exact gauge and supersymmetry
transformations preserving the
WZ gauge.
We also find that these transformations and 
the $O(C)$ action in the WZ gauge 
defined in the deformed harmonic
superspace reduces to the one in the ${\cal N}=1$ superspace by field
redefinition.

This paper is organised as follows:
In sect.\ 2, we review ${\cal N}=2$ supersymmetric $U(1)$ gauge theory in 
non(anti)commutative ${\cal N}=1$ superspace and
 non(anti)commutative ${\cal N}=2$ harmonic
superspace with generic non-singlet deformation parameters $C$.
For the latter superspace we construct 
 ${\cal N}=2$ supersymmetric $U(1)$ gauge theory in this
superspace at $O(C)$.
We then study the reduction of the deformation parameters in
non(anti)commutative  harmonic superspace to 
the deformed ${\cal N}=1$ superspace.
We find that the $O(C)$ action reduces to
the one defined in the deformed ${\cal N}=1$ superspace and argue  
invariance under ${\cal N}=(1,0)$ supersymmetry.
In sect.\ 3, 
we investigate the reduced theory in more detail.
We construct explicitly  the exact  gauge and ${\cal N}=(1,1/2)$ 
supersymmetry transformations
of the component fields in the WZ gauge, where detailed
calculations are explained in two appendices.
After the field redefinition, we find 
deformed ${\cal N}=(1,1/2)$ supersymmetry transformation  which keeps
the action in the deformed ${\cal N}=1$ superspace invariant.
Sect.\ 5 is devoted to discussion and conclusion.
In appendix \ref{app:UsefulFormulas}, 
we describe some useful reduction formulas of harmonic variables. 
The details of calculation of the 
exact ${\cal N}=(1,\, 1/2)$ supersymmetry transformation laws 
are presented in appendix \ref{app:DeterminationOfN=(1,1/2)}.

\section{${\cal N}=2$ supersymmetric $U(1)$ gauge theory \\in 
non(anti)commutative superspaces}
In this section we review ${\cal N}=2$ supersymmetric
$U(1)$ gauge theory in the deformed ${\cal N}=1$ superspace \cite{Se,ArItOh1}
and non(anti)commutative ${\cal N}=2$ harmonic superspace
\cite{IvLeZu,ArItOh3,ArItOh2}.

\subsection{The deformed ${\cal N}=1$ superspace}

Let $(x^{\mu},\theta^\alpha,\bar{\theta}^{\dot{\alpha}})$ be the
supercoordinates of ${\cal N}=1$ superspace\cite{WeBa}.
Here $\mu=0,1,2,3$ are spacetime indices,
$\alpha,\dot{\alpha}=1,2$ the spinor indices.
We study spacetime with the  Euclidean signature so that  chiral and
antichiral fermions transform independently.
We may call this ${\cal N}=1$ superspace as ${\cal N}=(1/2,1/2)$ superspace
as in \cite{IvLeZu}.
The non(anti)commutative ${\cal N}=1$ superspace 
is introduced by imposing
nonanticommutativity for Grassmann coordinates $\theta^\alpha$:
\begin{equation}
 \{ \theta^\alpha, \theta^\beta\}_{*}=C^{\alpha\beta},
\end{equation}
whereas the chiral coordinates
$y^{\mu}=x^{\mu}+i\theta\sigma^{\mu}\bar{\theta}$ commute with other 
coordinates  and 
$\bar{\theta}^{\dot{\alpha}}$ (anti)commute. 
The $*$-product $f*g$ is defined by $f*g=f\exp(P)g$,  where
$P=-{1\over2}\overleftarrow{Q}_{\alpha}C^{\alpha\beta}
\overrightarrow{Q}_{\beta}$ defines the Poisson
structure on the superspace.
Since this $Q$-deformation preserves chirality, {\it i.e.}
the Poisson structure $P$ commutes with the supercovariant derivatives
$D_{\alpha}={\partial \over \partial \theta^{\alpha}}
+i(\sigma^{\mu}\bar{\theta})_{\alpha}\partial_{\mu}$ and
$\bar{D}_{\dot{\alpha}}=-{\partial \over \partial
\bar{\theta}^{\dot{\alpha}}} -i(\theta
\sigma^{\mu})^{\dot{\alpha}}\partial_{\mu}$,
we can define
(anti)chiral superfields $\Phi$ ($\bar{\Phi}$) satisfying
$\bar{D}_{\dot{\alpha}}\Phi=0$ ($D_{\alpha}\bar{\Phi}=0$).
We can also introduce the vector superfield $V(y,\theta,\bar{\theta})$
for representing gauge fields.
If we take the WZ gauge for the vector superfield,
we are able to write down 
the Lagrangian in terms of component fields by replacing the product to
the $*$-product.
The component fields do not transform canonically under the gauge
transformation due to the deformation parameter $C$.
But in \cite{Se,ArItOh1}, the field redefinition was found
such that the component fields transform
canonically under the gauge transformation.

For ${\cal N}=2$ $U(1)$ gauge theory, matter fields $\Phi$, $\bar{\Phi}$ in the 
adjoint representation (under the $*$-product) 
and the vector superfield $V$
in the WZ gauge are given by 
\begin{eqnarray}
 \Phi(y,\theta)&=& A(y)+\sqrt{2}\theta\psi(y)+\bar{\theta}\bar{\theta}F(y),
\nonumber\\
\bar{\Phi}(\bar{y},\bar{\theta})&=&
\bar{A}(\bar{y})+\sqrt{2}\bar{\theta}\bar{\psi}(\bar{y})
+\bar{\theta}\bar{\theta}\left(
\bar{F}+2 i C^{\mu\nu} \partial_{\mu}(\bar{A}v_{\nu})
\right)(\bar{y}),
\nonumber\\
V(y,\theta,\bar{\theta})&=&
-\theta \sigma^{\mu}\bar{\theta} v_{\mu}(y)
+i \theta\theta \bar{\theta}\bar{\lambda}(y)
-i \bar{\theta}\bar{\theta}\theta^{\alpha}
\left(\lambda_\alpha
+{1\over2}\varepsilon_{\alpha\beta}C^{\beta\gamma}
(\sigma^{\mu}\bar{\lambda})_{\gamma} v_{\mu}
\right)(y)
\nonumber\\
&& +{1\over2} \theta\theta \bar{\theta}\bar{\theta}
(D-i\partial^{\mu}v_{\mu})(y),
\end{eqnarray}
where $\bar{y}^\mu=x^\mu-i \theta \sigma^\mu \bar{\theta}$ is the
anti-chiral coordinates.
The deformed action is defined by
\begin{equation}
S=\int d^4 xd^2\theta d^2\bar{\theta} \bar{\Phi}*e^{V}*\Phi*e^{-V}
+{1\over 16g^2}
\left(
\int d^2\theta W^{\alpha}*W_{\alpha}
+\int d^2\bar{\theta} \bar{W}_{\dot{\alpha}}*
\bar{W}^{\dot{\alpha}}
\right).
\label{eq:n12acrion1}
\end{equation}
Here $g$ denotes the gauge coupling constant.
The chiral and anti-chiral field strength are 
\begin{equation}
 W_{\alpha}=-{1\over4}\bar{D}\bar{D}e^{-V}D_{\alpha}e^{V},\quad
 \bar{W}_{\dot{\alpha}}={1\over4}DD e^{V}\bar{D}_{\dot{\alpha}}e^{-V}
\end{equation}
where  the multiplication of
superfields  is defined
by using the $*$-product.
After rescaling $V$ to $2gV$ and $C^{\alpha\beta}$ to ${1\over
2g}C^{\alpha\beta}$, 
this deformed action is expressed as
\begin{equation}
 S=\int d^4 x ({\cal L}^{(0)}+{\cal L}^{(1)})
\label{eq:n12action1}
\end{equation}
where
\begin{eqnarray}
{\cal L}^{(0)}&=&
-{1\over 4}v_{\mu\nu}(v^{\mu\nu}+\tilde{v}^{\mu\nu})
-i\bar{\lambda}\bar{\sigma}^{\mu}\partial_{\mu}\lambda
+{1\over2}D^2
-\partial^{\mu}\bar{A}\partial_{\mu}A-i\bar{\psi}\bar{\sigma}^{\mu}\partial_{\mu}\psi
+\bar{F}F,
\label{eq:n12componetaction0}\\
{\cal L}^{(1)}&=& -{i\over2} C^{\mu\nu}v_{\mu\nu} (\bar{\lambda}\bar{\lambda})
+\sqrt{2} C^{\alpha\beta}\psi_\alpha (\sigma^{\mu}\bar{\lambda})_{\beta} 
\partial_{\mu}\bar{A} +i C^{\mu\nu} v_{\mu\nu} \bar{A}F,
\label{eq:n12componetaction1}
\end{eqnarray}
and $v_{\mu\nu}=\partial_{\mu}v_{\nu}-\partial_{\nu}v_{\mu}$.
This action is invariant under the gauge transformation 
$\delta_{\lambda}v_{\mu}=-\partial_{\mu}\lambda$.
Undeformed theory has ${\cal N}=2$ 
extended supersymmetry, where only ${\cal N}=1$ symmetry generated by
$Q_\alpha$ and $\bar{Q}_{\dot{\alpha}}$ is manifest in
${\cal N}=1$ superspace formalism. 
In deformed theory, 
the generator $\bar{Q}_{\dot{\alpha}}$ does not commute with the 
Poisson structure $P$ and is not a symmetry of the theory.
On the other hand, $Q_{\alpha}$ generates a symmetry of the theory.
Since $Q$ transformation does not preserve the 
WZ gauge, we need to do gauge transformation to retain the WZ gauge.
The deformed supersymmetry transformation is 
\begin{eqnarray}
&& 
\delta^*_{\xi}A
= 
	 \sqrt{2}\xi\psi
	,\qquad
\delta^*_{\xi}\bar{A}
=
	0
	,\nonumber\\
&& 
\delta^*_{\xi} \psi_{\alpha} 
= 
	\sqrt{2} \xi_{\alpha} F
	, \qquad 
\delta^*_{\xi} \bar{\psi}_{\dot{\alpha}} 
= 
	-i\sqrt{2} (\xi \sigma^{\mu})_{\dot{\alpha}} 
    \partial_{\mu} \bar{A}
	, \nonumber\\ 
&& 
\delta^*_{\xi} F 
= 
	0
	, \qquad 
\delta^*_{\xi} \bar{F} 
= 
	i\sqrt{2} (\xi \sigma^{\mu} \partial_{\mu} \bar{\psi}) 
   - 2 C^{\alpha\beta} \xi_\alpha \sigma^{\mu}_{\beta\dot{\alpha}} 
    \partial_{\mu} (\bar{\lambda}^{\dot{\alpha}} \bar{A})
	, \nonumber\\  
&& 
\delta^*_{\xi}v_{\mu}
= 
	i \xi\sigma_\mu\bar{\lambda}
	,\nonumber\\ 
&& 
\delta^*_{\xi} \lambda_{\alpha} 
= 
	(\sigma^{\mu\nu} \xi)_{\alpha} 
    \left\{ 
      v_{\mu\nu} 
     +\frac{i}{2} C_{\mu\nu} (\bar{\lambda} \bar{\lambda}) 
    \right\} 
   +i \xi_{\alpha} D 
	, \qquad 
\delta^*_{\xi} \bar{\lambda}_{\dot{\alpha}} 
= 
	0,                                               \nonumber\\ 
&& 
\delta^*_{\xi} D 
= 
	-(\xi \sigma^{\mu} \partial_{\mu} \bar{\lambda}). 
\label{eq:n12susytr1}
\end{eqnarray}
Remaining ${\cal N}=1$ supersymmetry, however, is not manifest
in this formalism.
In order to obtain this supersymmetry transformation more
systematically,
it is convenient to introduce non(anti)commutative extended superspace.
In the next section,  we will 
discuss ${\cal N}=2$ supersymmetric $U(1)$ gauge theory in
non(anti)commutative harmonic superspace.

\subsection{Non(anti)commutative ${\cal N}=2$  harmonic superspace}
Next  
we review non(anti)commutative deformation of 
${\cal N}=2$ harmonic superspace and ${\cal N}=2$ supersymmetric $U(1)$ gauge 
theory in this superspace\cite{ArItOh3,ArItOh2}.
Let ($x^{\mu},\theta_\alpha^i, \bar{\theta}_{\dot{\alpha}}^i$) be 
the coordinates of ${\cal N}=2$ (rigid) superspace.
The index $i=1,2$ labels
the doublet of the $SU(2)_R$ $R$-symmetry.
The supersymmetry generators $Q^{i}_{\alpha}$,
$\bar{Q}_{\dot{\alpha}i}$ and the supercovariant
derivatives $D_{\alpha}^{i}$, $\bar{D}_{\dot{\alpha}i}$ are
defined by
\begin{eqnarray}
Q_{\alpha}^{i}&=&{\partial\over\partial\theta^{\alpha}_{i}}
-i(\sigma^{\mu})_{\alpha\dot{\alpha}}\bar{\theta}^{\dot{\alpha} i}
{\partial\over\partial x^{\mu}}, \quad
\bar{Q}_{\dot{\alpha} i}= -{\partial\over\partial\bar{\theta}^{\dot{\alpha} i}}
+i\theta^{\alpha}_{i}(\sigma^{\mu})_{\alpha\dot{\alpha}}
{\partial\over \partial x^{\mu}}
, 
\nonumber\\
D_{\alpha}^{i}&=&{\partial\over\partial\theta^{\alpha}_{i}}
+i(\sigma^{\mu})_{\alpha\dot{\alpha}}\bar{\theta}^{\dot{\alpha} i}
{\partial\over\partial x^{\mu}}, \quad
\bar{D}_{\dot{\alpha} i}= -{\partial\over\partial\bar{\theta}^{\dot{\alpha} i}}
-i\theta^{\alpha}_{i}(\sigma^{\mu})_{\alpha\dot{\alpha}}
{\partial\over \partial x^{\mu}}.
\end{eqnarray}
The ${\cal N}=2$ harmonic superspace \cite{GaIvOgSo} is introduced by adding 
the harmonic variables $u^{\pm}_i$ to the ${\cal N}=2$ superspace coordinates.
The variables $u^{\pm}_i$ form an $SU(2)$ matrix and satisfy the conditions
$ u^{+i}u^{-}_{i}=1$ and $\overline{u^{+i}}=u^{-}_{i}$.
The completeness condition for $u^{\pm}_i$ reads
$u^{+}_{i}u^{-}_{j}-u^{+}_{j}u^{-}_{i}=\epsilon_{ij}$.
Using $u^{\pm}_{i}$, the $SU(2)_R$ indices can be projected into two
parts with $\pm 1$ $U(1)(\subset SU(2)_R)$ charges.
For example, we define the supercovariant derivatives 
$D^{\pm}_{\alpha}$ and $\bar{D}^{\pm}_{\alpha}$ by
$
 D^{\pm}_{\alpha}=u^{\pm}_{i}D^{i}_{\alpha}$, 
$\bar{D}^{\pm}_{\alpha}=u^{\pm}_{i}\bar{D}^{i}_{\alpha}
$.
$D^{i}_{\alpha}$ is solved by $D^{\pm}_{\alpha}$ such as
$D^{\pm}_{\alpha}=u^{+}_{i}D^{-}_{\alpha}-u^{-}_{i}D^{+}_{\alpha}$
with the help of the completeness condition.
In the harmonic superspace formalism, an important ingredient is an
analytic superfield rather than the ${\cal N}=2$ chiral superfield.
An analytic superfield $\Phi(x,\theta,\bar{\theta},u)$ is defined by
$D^{+}_{\alpha}\Phi=\bar{D}^{+}_{\dot{\alpha}}\Phi=0$.
It is convenient to write this analytic superfield in terms of analytic
basis:
$
 x_{A}^{\mu}= 
x^{\mu}-i (\theta^i \sigma^\mu \bar{\theta}^j +\theta^j
\sigma^\mu\bar{\theta}^i)
u^{+}_{i}u^{-}_{j}$,
$\theta^{\pm}_{\alpha}= u^{\pm }_{i}\theta^{i}_{\alpha}$ and 
$\bar{\theta}^{\pm}_{\dot{\alpha}}= u^{\pm}_{i}\bar{\theta}^{i}_{\dot{\alpha}}$.
In this basis, 
an analytic superfield $\Phi$ is functions
of $(x^{\mu}_{A},\theta^{+},\bar{\theta}^{+},u)$: 
$
 \Phi=\Phi(x^{\mu}_{A},\theta^{+},\bar{\theta}^{+},u)
$. 

We now introduce the nonanticommutativity in the ${\cal N}=2$ harmonic
superspace by using the $*$-product:
\begin{equation}
 f*g(\theta)=f(\theta)\exp(P) g(\theta),\quad
P=-{1\over2}
\overleftarrow{Q^{i}_{\alpha}}
C^{\alpha\beta}_{ij}\overrightarrow{Q^{j}_{\beta}}, 
\label{eq:moyal2}
\end{equation}
where $C^{\alpha\beta}_{ij}$ is some constants. 
With this $*$-product, we have following (anti)commutation relations: 
\begin{equation}
 \{ \theta^{\alpha}_{i}, \theta^{\beta}_{j}\}_{*}=C^{\alpha\beta}_{ij}
, \quad 
 [x^{\mu}_{L}, x^{\nu}_{L}]_{*}=[x^{\mu}_{L}, \theta^{\alpha}_{i}]_{*}=
[x^{\mu}_{L},\bar{\theta}^{\dot{\alpha} i}]_{*}=0, \quad 
\{ \bar{\theta}^{\dot{\alpha} i}, \bar{\theta}^{\dot{\beta} j}\}_{*}=
\{ \bar{\theta}^{\dot{\alpha} i}, \theta^{\alpha}_{j}\}_{*}=0
, 
\end{equation}
where $x_{L}^{\mu}\equiv x^\mu+i\theta_i \sigma^\mu \bar{\theta}^i$ are 
${\cal N}=2$ chiral coordinates. 
The deformation parameter $C^{\alpha\beta}_{ij}$ is symmetric
under the exchange of pairs of indices $(\alpha i)$,$(\beta j)$:
 $C^{\alpha\beta}_{ij}=C^{\beta\alpha}_{ji}$.
We decompose the nonanticommutative parameter $C^{\alpha\beta}_{ij}$
into the symmetric and antisymmetric parts with respect to the $SU(2)$ indices,
such as
\begin{equation}
C^{\alpha\beta}_{ij}=C^{\alpha\beta}_{(ij)}
+{1\over4}\epsilon_{ij}\varepsilon^{\alpha\beta}C_{s}.
\end{equation}
Here we denote $A_{(i_1\cdots i_n)}$ by the symmetrized sum of
$A_{i_1\cdots i_n}$ over indices $i_1,\cdots, i_n$.
$C^{\alpha\beta}_{ij}$ with zero $C^{\alpha\beta}_{(ij)}$ 
corresponds to the singlet deformation \cite{IvLeZu,ArIt3,FeIvLeSoZu}.
Since $P$ commutes with the supercovariant derivatives 
$D$, the chiral structure is preserved by this deformation.
Since 
we will consider the non-singlet deformation in the following, we put
$C_s=0$.

We now construct the action of ${\cal N}=2$ supersymmetric $U(1)$
gauge theory in this non(anti)commutative superspace.
We introduce an analytic superfield $V^{++}(\zeta,u)$ 
with $\zeta=(x_A^{\mu},\theta^+, \bar{\theta}^+)$ by covariantizing
the harmonic derivative 
$D^{++}
= 
        u^{+i}{\partial\over \partial u^{-i}}
        - 2i \theta^+ \sigma^\mu \bar{\theta}^+ 
                {\partial\over \partial x_A^\mu } 
        + \theta ^{+\alpha} {\partial \over \partial \theta^{-\alpha}} 
        + \bar{\theta}^{+\dot{\alpha}} 
                {\partial \over \partial \bar{\theta}^{-\dot{\alpha}}}
\rightarrow \nabla^{++}=D^{++}+i V^{++}$.
Generalizing the construction in \cite{Zu,BuSa}, the action is given by
\begin{equation}
S_{*}
= 
\frac12 \sum_{n=2}^{\infty} \int d^4xd^8\theta du_1\dots du_n {(-i)^n \over n}
{ V^{++}(1) * \cdots * V^{++}(n) 
\over (u_1^+ u_2^+)\cdots (u_{n}^+ u_1^+) }
,  
\label{eq:StarDeformedAction:Gen}
\end{equation} 
where $V^{++}(i)=V^{++}(\zeta_i, u_i)$,
$\zeta_{i}=(x_{A},\theta^{+}_{i},\bar{\theta}^{+}_{i})$ and 
$d^8\theta=d^4 \theta^+ d^4\theta^- $ with
$d^4\theta^{\pm}=d^2\theta^{\pm}d^2\bar{\theta}^{\pm}$.
The harmonic integral is defined by the rules:
(i)
$
 \int du f(u)=0
$
for $f(u)$ with non-zero $U(1)$ charge.
 (ii)
$
 \int du 1=1.
$
(iii)
$
 \int du u^{+}_{(i_1}\cdots u^{+}_{i_n} u^{-}_{j_1}\cdots u^{-}_{j_n)}=0,
\ (n\geq 1).
$
The action (\ref{eq:StarDeformedAction:Gen}) is invariant under the 
gauge transformation
\begin{equation}
 \delta_\Lambda^{*} V^{++}=-D^{++}\Lambda+i [ \Lambda, V^{++}]_{*}
, 
\label{eq:gauge2}
\end{equation}
with an analytic superfield $\Lambda$.
The generic superfield $V^{++}(\zeta,u)$ includes infinitely many
auxiliary fields. 
Most of these fields are gauged away except the lowest component fields 
in the harmonic expansion. 
One can take the WZ gauge
\begin{eqnarray}
 V^{++}_{WZ}(x_{A},\theta^{+},\bar{\theta}^{+},u)
&=& 
-i\sqrt{2}(\theta^{+})^2 \bar{\phi}(x_{A})
+i\sqrt{2}(\bar{\theta}^{+})^2 \phi(x_{A})
-2i \theta^{+}\sigma^{\mu}\bar{\theta}^{+}A_{\mu}(x_{A})\nonumber\\
&&+4(\bar{\theta}^{+})^2\theta^{+}\psi^{i}(x_{A}) u^{-}_{i}
-4(\theta^{+})^2\bar{\theta}^{+}\bar{\psi}^{i}(x_{A})u^{-}_{i}\nonumber\\
&&
+3(\theta^{+})^2(\bar{\theta}^{+})^2 D^{ij}(x_{A})u^{-}_{i} u^{-}_{j},
\label{eq:wz1}
\end{eqnarray}
which is convenient to study the theory in the component formalism. 

The component action $S_*$ in the WZ gauge can be expanded in a power series
of the deformation parameter $C$. 
In \cite{ArItOh2}, we have computed the $O(C)$ action explicitly.
The quadratic part $S_{*,2}$ in $S_{*}$ is the same as the commutative
one:
\begin{equation}
S_{*,2}=\int d^4 x\left\{
-{1\over 4}F_{\mu\nu}F^{\mu\nu}-{1\over4}F_{\mu\nu}\tilde{F}^{\mu\nu}
+\phi \partial^2 \bar{\phi}
-i\psi^{i}\sigma^{\mu}\partial_{\mu}\bar{\psi}_{i}
+{1\over4}D^{ij}D_{ij}
\right\}.
\label{eq:abelianaction}
\end{equation}
The cubic part $S_{*,3}$ in $S_{*}$ is of order $O(C)$ and given by
\begin{eqnarray}
S_{*,3}
&=&{}
        \int d^4x \left[
- { 2 \sqrt{2} \over 3} i C_{(ij)}^{\alpha\beta}
        \psi^i_\alpha ( \sigma^\nu \partial_\nu \bar{\psi}^j)_\beta 
        \bar{\phi}
- 2\sqrt{2} i C_{(ij)}^{\alpha\beta}
        \psi^i_\alpha (\sigma^\nu \bar{\psi}^j)_\beta 
        \partial_\nu \bar{\phi}
        \right. \nonumber\\ 
&&{}
+{2\over 3} i C_{(ij)}^{\alpha\beta} A_\mu
        (\sigma^\mu \bar{\psi}^i)_\alpha (\sigma^\nu \partial_\nu \bar{\psi}^j)_\beta
- i C_{(ij)}^{\mu\nu} \bar{\psi}^i \bar{\psi}^j  F_{\mu\nu}
\nonumber\\ 
&&\left. {} 
+ \sqrt{2} C_{(ij)}^{\mu\nu} D^{ij} A_\mu \partial_\nu \bar{\phi}
+ {1\over \sqrt{2}} C_{(ij)}^{\mu\nu} D^{ij} F_{\mu\nu} \bar{\phi} 
\right] 
. 
\label{eq:thirdorderLagrangian}
\end{eqnarray}
Note that here we have already dropped the $C_s$ dependent terms. 
We will refer 
$
S_{*,2}+ S_{*,3} 
$
as the $O(C)$ action. 

In the commutative case, the gauge parameter $\Lambda=\chi(x_A)$
preserves the WZ gauge and gives rise to the gauge transformation for
component fields.
In the non(anti)commutative case, however, 
the gauge transformation (\ref{eq:gauge2}) with the same gauge
parameter does not preserve the WZ gauge because of the $C$-dependent
terms arising from the commutator. 
In order to preserve the WZ gauge, one must include the $C$-dependent 
terms. The gauge parameter
is shown to take the form 
\begin{eqnarray}
\lambda_C (\zeta, u)
&=&
        \chi(x_A)
        + \theta^{+}\!\sigma^\mu \bar{\theta}^{+}
                \lambda_\mu^{(-2)} (x_A, u; C)
        + (\bar{\theta}^{+})^2 
                \lambda^{(-2)} (x_A, u; C)
                \nonumber\\
&&{} 
        + (\bar{\theta}^{+})^2 \theta^{+}{}^{\alpha} 
                \lambda_{\alpha}^{(-3)} (x_A, u; C)
        + (\theta^{+})^2 (\bar{\theta}^{+})^2 
                \lambda^{(-4)} (x_A, u; C)
, 
\label{eq:gaugeparam1}
\end{eqnarray}
which has been determined by solving the 
WZ gauge preserving conditions expanded  in harmonic modes
\cite{ArItOh2}. 
The gauge transformation is also fully determined, which reads 
\begin{eqnarray}
\delta^{*}_{\lambda_C} A_\mu 
&=& 
        - \partial_\mu \chi 
        + O(C^2),
\nonumber\\
\delta^{*}_{\lambda_C} \phi 
&=&
        O(C^2),
\nonumber\\
\delta^{*}_{\lambda_C} \psi_{\alpha i} 
&=& 
        \frac23 (\varepsilon C_{(ij)} \sigma^\mu \bar{\psi}^j )_\alpha 
        \, \partial_{\mu} \chi 
        +O(C^2) ,
\nonumber\\
\delta^{*}_{\lambda_C} D_{ij}
&=&
        2 \sqrt{2} C_{(ij)}^{\mu\nu} \partial_\mu \chi \partial_\nu \bar{\phi} 
        +O(C^2) 
	, \nonumber\\
\delta_{\lambda_C}^{*} (\mbox{others}) 
&=& 
        0
        . 
\label{eq:gaugetr1}
\end{eqnarray}
The $O(C)$ action is 
invariant under the $O(C)$ gauge transformation (\ref{eq:gaugetr1}).

These gauge transformations are not canonical.
But if we redefine the component fields such as
\begin{eqnarray}
\hat{A}_\mu 
&=& 
        A_\mu 
        + O(C^2) 
        ,\nonumber\\ 
\hat{\phi} 
&=& 
        \phi 
        + O(C^2)
        , 
\quad \hat{\bar{\phi}}=\bar{\phi},
\nonumber\\ 
\hat{\psi}_{\alpha i} 
&=& 
        \psi_{\alpha i} 
        +\frac23 (\varepsilon C_{(ij)} \sigma^\mu \bar{\psi}^j )_\alpha A_{\mu} 
        +O(C^2) 
        , \quad
\hat{\bar{\psi}}{}^{\dot{\alpha}}=\bar{\psi}^{\dot{\alpha}}
, \nonumber\\
\hat{D}_{ij} 
&=&  
        D_{ij} 
        + 2 \sqrt{2} C_{(ij)}^{\mu\nu} A_\mu \partial_\nu \bar{\phi} 
        +O(C^2),
\label{eq:orderCgaugetr}
\end{eqnarray}
the newly defined fields are shown to transform canonically: $
\delta_{\lambda_C}^* \hat{A}_{\mu}=-\partial_{\mu}\chi$, 
$\delta_{\lambda_C}^* (\mbox{others})=0$.
In terms of redefined fields, the $O(C)$ action can be written as
\begin{eqnarray}
S_{*,2} + S_{*,3} 
&=& 
        \int d^4x \left[
        - {1\over 4} 
                \hat{F}_{\mu\nu} ( \hat{F}^{\mu\nu} + \tilde{\hat{F}}{}^{\mu\nu} ) 
        + \hat{\phi} \partial^2 \hat{\bar{\phi}}
        -i \hat{\psi}^i \sigma^\mu \partial_\mu \hat{\bar{\psi}}_i
        + \frac14 
                 \hat{D}^{ij} \hat{D}_{ij}
                \right. \nonumber\\
&&{}
        - 2\sqrt{2} i C_{(ij)}^{\alpha\beta}
                \hat{\psi}^i_\alpha (\sigma^\mu \hat{\bar{\psi}}{}^j)_\beta 
                \partial_\mu \hat{\bar{\phi}}
        - { 2 \sqrt{2} \over 3} i C_{(ij)}^{\alpha\beta}
        \hat{\psi}^i_\alpha ( \sigma^\mu \partial_\mu 
\hat{\bar{\psi}}{}^j)_\beta 
                \hat{\bar{\phi}}
        \nonumber\\
&&\left.{}
        - i C_{(ij)}^{\mu\nu} \hat{\bar{\psi}}{}^i \hat{\bar{\psi}}{}^j 
\hat{F}_{\mu\nu}
        + {1\over \sqrt{2}} C_{(ij)}^{\mu\nu} \hat{D}^{ij} \hat{F}_{\mu\nu}
\hat{\bar{\phi} }
        + O(C^2)
        \right] 
,
\label{eq:redefinedaction}
\end{eqnarray}
where $\hat{F}_{\mu\nu}=\partial_{\mu}\hat{A}_{\nu}
-\partial_{\nu}\hat{A}_{\mu}$.

The $O(C)$ deformed ${\cal N}=2$ supersymmetry transformation of the component
 fields in the WZ gauge is determined in a similar way\cite{ArItOh3}.
After the field redefinition (\ref{eq:orderCgaugetr}), the transformation is 
given by 
\begin{eqnarray}
\delta^{*}_{\xi} \hat{\phi} 
&=& 
        - \sqrt{2} i \xi^{i} \hat{\psi}_{i}
        - \frac83 i ( \xi^{j} \varepsilon C_{(j k)} \hat{\psi}^{k} ) \hat{\bar{\phi}} 
        + O(C^2)
        , \nonumber \\
\delta^{*}_{\xi} \hat{\bar{\phi}}
&=& 
        0 
        , \nonumber \\
\delta^{*}_{\xi} \hat{A}_\mu 
&=& 
        i \xi^{i} \sigma_\mu \hat{\bar{\psi}}{}_{i}
        + 2 \sqrt{2}i ( \xi^{j} \varepsilon C_{(j k)} \sigma_\mu \hat{\bar{\psi}
}{}^{k} ) 
                \hat{\bar{\phi}}
        + O(C^2)
        , \nonumber \\
\delta^{*}_{\xi} \hat{\psi}^{\alpha i} 
&=& 
        - (\xi^{i} \sigma^{\mu\nu} )^{\alpha} \hat{F}_{\mu\nu}   
        - \hat{D}^{ij} \xi^\alpha_j 
        - i ( \xi^{i} \sigma_{\mu\nu} )^\alpha 
                C_{(j k)}^{\mu\nu} (\hat{\bar{\psi}}{}^{j} \hat{\bar{\psi}}{}^{k
}) 
        + 2 \sqrt{2} \hat{D}^{(i j} ( \xi^{k)} \varepsilon C_{(j k)} )^\alpha 
                \hat{\bar{\phi}} 
                        \nonumber\\
&&{} 
        - \left\{  
                2 \sqrt{2} (\xi^{j} \varepsilon C_{(j k)} 
                        \sigma^{\mu\nu} )^\alpha 
                + {2\sqrt{2}\over 3} (\xi^{j} \sigma^{\mu\nu} \varepsilon 
                        C_{(j k)} )^\alpha 
                + \sqrt{2} C_{(j k)}^{\mu\nu} \xi^{\alpha j} 
        \right\} 
                \epsilon^{k i} \hat{\bar{\phi}} \hat{F}_{\mu\nu} 
        + O(C^2)
        , \nonumber \\ 
\delta^{*}_{\xi} \hat{\bar{\psi}}{}^{i}_{\dot{\alpha}}  
&=& 
        + \sqrt{2} (\xi^{i} \sigma^\nu )_{\dot{\alpha}} 
                \partial_\nu \hat{\bar{\phi}} 
        + 2 
                (\xi^{j} \varepsilon C_{(j k)} \sigma^\nu)_{\dot{\alpha}} 
                \partial_\nu (\hat{\bar{\phi}}{}^2) 
                \epsilon^{k i}
        + O(C^2) 
        , \nonumber \\
\delta^{*}_{\xi} \hat{D}^{ij} 
&=& 
        - 2 i \xi^{(i} \sigma^\nu \partial_\nu \hat{\bar{\psi}}{}^{j)}
                        \nonumber\\
&&{} 
        - 6 \sqrt{2} i 
                \epsilon^{k(l} \partial_\nu \bigl\{
                (\xi^{i} \varepsilon C_{(k l)} \sigma^\nu 
                \hat{\bar{\psi}}{}^{j)}) \hat{\bar{\phi}} \bigr\} 
        + 2\sqrt{2} i \epsilon^{il}\epsilon^{jm} 
                        ( \xi^k \varepsilon C_{(lm)} \sigma^\nu \hat{\bar{\psi}}
{}_k ) 
                \partial_\nu \hat{\bar{\phi}} 
        + O(C^2)
. 
\label{eq:DeformedSUSYTr2}
\end{eqnarray} 

\subsection{Reduction to the deformed ${\cal N}=1$ superspace}
Since the deformed ${\cal N}=2$ action (\ref{eq:n12action1}) 
in the deformed ${\cal N}=1$ superspace
contains only $O(C)$ corrections,
it is important to compare ${\cal N}=1$ superspace formalism to
non(anti)commutative harmonic superspace formalism.
We will examine this correspondence in terms of  component 
fields because it is difficult to find the relationship 
between ($\Phi$,$\bar{\Phi}$,$V$) and $V^{++}_{WZ}$ explicitly at the
superfield level.

We impose the condition for the deformation parameters
$C^{ij}_{\alpha\beta}$ such as
\begin{equation}
 C^{\alpha\beta}_{ij} =C_{11}^{\alpha\beta}\delta_{i}^{1}\delta_{j}^{1}.
\label{eq:reduced1}
\end{equation}
This reduction of the parameters implies that only the $\theta^i_\alpha$ 
coordinates are nonanticommutative and $\theta^2_\alpha$ are ordinary Grassmann 
coordinates.
Then the action (\ref{eq:redefinedaction})
 with redefined component fields becomes
\begin{eqnarray}
S_{*,2}+S_{*,3}
&=&
        \int d^4x \left[
        - {1\over 4} \hat{F}_{\mu\nu} ( \hat{F}^{\mu\nu} 
+ \tilde{\hat{F}}{}^{\mu\nu} )
        -i \hat{\psi}^i \sigma^\mu \partial_\mu \hat{\bar{\psi}}_i
        + \hat{\phi} \partial^{\mu}\partial_{\mu} \hat{\bar{\phi}}
        + \frac14 \hat{D}_{ij} \hat{D}^{ij}
        \right. 
        \nonumber\\
&&{} 
- { 2 \sqrt{2} \over 3} i C_{11}^{\alpha\beta}
        \hat{\psi}^1_\alpha ( \sigma^\nu \partial_\nu \hat{\bar{\psi}}{}^1)_\beta 
        \hat{\bar{\phi}}
- 2\sqrt{2} i C_{11}^{\alpha\beta}
        \hat{\psi}^1_\alpha (\sigma^\nu \hat{\bar{\psi}}{}^1)_\beta 
        \partial_\nu \hat{\bar{\phi}}
        \nonumber\\
&&\left. {} 
- i C_{11}^{\mu\nu} \hat{\bar{\psi}}{}^1 \hat{\bar{\psi}}{}^1  \hat{F}_{\mu\nu}
+ {1\over \sqrt{2}} C_{11}^{\mu\nu} \hat{D}^{11} \hat{F}_{\mu\nu} 
\hat{\bar{\phi}} 
\right] 
. 
\label{eq:redefinedaction2}
\end{eqnarray}
Since both component fields $(\hat{\phi},\hat{\bar{\phi}}, \hat{\psi}^i, 
\hat{\bar{\psi}}{}^i,\hat{D}^{ij})$
and
$(A,\bar{A},\psi,\bar{\psi},\lambda, \bar{\lambda},v_{\mu},F,\bar{F},D)$ 
transform canonically under the gauge
transformation,
these fields must be related to each other in gauge invariant way.
By comparing (\ref{eq:n12action1}) and (\ref{eq:redefinedaction2}),
we find 
\begin{eqnarray}
v_{\mu}&=&\hat{A}_{\mu}, \quad
A= -i\hat{\phi}, \quad \bar{A}=i\hat{\bar{\phi}}, \nonumber\\
D&=& i\hat{D}^{12},\quad F=-{1\over\sqrt{2}}\hat{D}^{11}, \quad \bar{F}=
-{1\over\sqrt{2}}(\hat{D}^{22}
-\sqrt{2} C_{11}^{\mu\nu} \hat{F}_{\mu\nu} \hat{\bar{\phi}}), 
\nonumber\\
\lambda^\beta&=&\hat{\psi}^{2\beta}-{2\sqrt{2}\over3} 
C_{11}^{\alpha\beta}\hat{\psi}^1_\alpha \hat{\bar{\phi}},
  \quad \psi=\hat{\psi}^1,\quad
\bar{\psi}=\hat{\bar{\psi}}_1,\quad \bar{\lambda}=\hat{\bar{\psi}}_2.
\label{eq:identification2}
\end{eqnarray}
where the deformation parameter $C_{11}^{\alpha\beta}$ is related to
$C^{\alpha\beta}$ by
\begin{equation}
 C_{11}^{\alpha\beta}={1\over 2} C^{\alpha\beta}.
\label{eq:identification3}
\end{equation}
Note that for $C=0$ we get the field redefinition for undeformed theory.

We now consider the deformed supersymmetry transformations.
Taking the Grassmann parameters as $(\xi^1,\xi^2)=(-\eta,\xi)$ and
using the supersymmetry transformations (\ref{eq:DeformedSUSYTr2}) and 
identifications
(\ref{eq:identification2}), it is shown that the transformation
associated with $\xi$ becomes (\ref{eq:n12susytr1}).
The $Q^2$ supersymmetry transformation associated with the parameter $\eta$
would give a remaining transformation.
But it turns out that $O(C)$ transformation with the same identification 
does not keep the action invariant.
It is necessary to introduce $O(C^2)$ correction to the $Q^2$
supersymmetry transformation.
We found that the following transformation keep the action invariant: 
\begin{eqnarray} 
&& 
\delta_{\eta}^* A 
= 
	\sqrt{2} \eta \lambda 
	- 2 i \eta \varepsilon C \psi \bar{A} 
	, 
	\qquad 
\delta_{\eta}^* \bar{A} 
= 
	0 
	, 
	\nonumber\\ 
&& 
\delta_{\eta}^* \psi^\alpha 
= 
	( \eta \sigma_{\mu\nu} )^\alpha 
		\left\{ 
		v^{\mu\nu} 
		+ \frac i2 C^{\mu\nu} ( \bar{\lambda}\bar{\lambda} - 2 F \bar{A} ) 
		\right\} 
	+ i \eta^\alpha D 
	, 
	\qquad 
\delta_{\eta}^* \bar{\psi}_{\dot{\alpha}} 
= 
	- ( \eta \varepsilon C \sigma^\nu )_{\dot{\alpha}} 
		\partial_\nu ( \bar{A}^2 ) 
	, 
	\nonumber\\ 
&& 
\delta_{\eta}^* F 
= 
	\sqrt{2} i \eta \sigma^\nu \partial_\nu \bar{\lambda} 
	, 
	\qquad 
\delta_{\eta}^* \bar{F} 
= 
	2 \sqrt{2} i \det C \eta \sigma^\nu 
		\partial_\nu ( \bar{\lambda} \bar{A}^2 ) 
	, 
	\nonumber\\ 
&& 
\delta_{\eta}^* v_\mu 
= 
	- i \eta \sigma_\mu \bar{\psi} 
	+ \sqrt{2} ( \eta \varepsilon C \sigma_\mu \bar{\lambda} ) \bar{A}
	, 
	\nonumber\\	
&& 
\delta_{\eta}^* \lambda^\alpha 
= 
	\sqrt{2} \eta^\alpha \bar{F} 
	+ \sqrt{2} ( \eta \varepsilon C )^\alpha \bar{A} D 
	- 
	\sqrt{2} 
	i ( \eta \sigma^{\mu\nu} \varepsilon C )^\alpha 
		v_{\mu\nu} \bar{A} 
	+ 
	\sqrt{2} 
	\det C \eta^\alpha \bar{\lambda}\bar{\lambda} \bar{A} 
	, 
	\nonumber\\ 
&& 
\delta_{\eta}^* \bar{\lambda}_{\dot{\alpha}} 
= 
	- \sqrt{2} i ( \eta \sigma^\nu )_{\dot{\alpha}} \partial_\nu \bar{A} 
	, 
	\nonumber\\ 
&& 
\delta_{\eta}^* D 
= 
	- \eta \sigma^\nu \partial_\nu \bar{\psi} 
	+ \sqrt{2} i \eta \varepsilon C \sigma^\nu 
		\partial_\nu ( \bar{\lambda} \bar{A} ) 
	, 
\label{eq:etasusy1}
\end{eqnarray}
where 
$(\varepsilon C)_\alpha{}^\beta \equiv \varepsilon_{\alpha\gamma} C^{\gamma\beta}$. 
Hence we have found the 
${\cal N}=(1,0)$ supersymmetry of the ${\cal N}=2$ action
in the deformed ${\cal N}=1$ superspace.

In (\ref{eq:etasusy1}), we add $O(C^2)$ correction to the transformation
 by hand.
This correction is not unique because the action (\ref{eq:n12action1}) 
is invariant under the
transformation
$\tilde{\delta}_\eta$ with
\begin{eqnarray}
 \tilde{\delta}_\eta \lambda_{\alpha}&=&\eta_\alpha f_{1}(\bar{A})F,
\nonumber\\
 \tilde{\delta}_\eta \bar{F}&=& i f_{1}(\bar{A}) 
(\eta \sigma^{\mu}\partial_{\mu}\bar{\lambda}),
\end{eqnarray}
for arbitrary function $f_{1}(\bar{A})$ of $\bar{A}$.
In the next section, we will show that we are
able to construct exact supersymmetry transformation for 
the reduced deformation parameter (\ref{eq:reduced1}).
If we use the field redefinition (\ref{eq:identification2}), we can fix
$f_1(\bar{A})=0$.
The result is shown to be exactly equal to (\ref{eq:etasusy1}).


We note also that $\bar{Q}_2$ supersymmetry transformation commutes 
with the Poisson structure $P$.
The theory is expected to have ${\cal N}=(1,1/2)$ extended supersymmetry
\cite{IvLeZu}.
In the next section we will construct the exact ${\cal N}=(1,1/2)$ 
supersymmetry transformation in the framework of the harmonic superspace formalism.

\section{${\cal N}=(1,\, 1/2)$ Supersymmetry}
\label{sect:N=(1,1/2)SUSY}
In this section, 
we will present the ${\cal N}=(1,\, 1/2)$ 
supersymmetry transformation 
generated by $Q^1$, $Q^2$ and $\bar{Q}_2$, 
within the harmonic superspace formalism. 
Under the restriction of the deformation parameter (\ref{eq:reduced1}),  
we can determine 
the ${\cal N}=(1,\, 1/2)$ supersymmetry transformation laws exactly. 
The main concern is the contribution from the deformed gauge transformation 
to retain the WZ gauge. 
Although the calculation is much simpler than 
the case of the generic deformation parameter, 
it turns out to be considerably lengthy 
even under the restriction. 
On the other hand, one finds that the determination of 
the exact gauge transformation laws is accomplished 
in a similar way 
but is much easier than the supersymmetry. 
Therefore, in order to illustrate how to determine the exact transformations, 
first we derive in sect.\ \ref{sect:ExactGaugeTr} 
the exact gauge transformation laws in \cite{ArItOh2}, 
restricting ourselves to 
the reduced deformation parameter (\ref{eq:reduced1}). 
The exact ${\cal N}=(1,\, 1/2)$ supersymmetry transformation 
is given in sect.\ \ref{sect:ExactN=(1,1/2)}, 
though the details of the actual calculation 
are presented in appendix \ref{app:DeterminationOfN=(1,1/2)}. 
From the exact gauge transformation, 
we can find a field redefinition that 
leads to the canonical component gauge transformation.  
We also give the ${\cal N}=(1,\, 1/2)$ transformation 
after this field redefinition.

\subsection{Determination of the exact transformation laws}
\label{sect:ExactGaugeTr} 

In this subsection, 
in order to demonstrate how to determine  
the exact transformation laws 
under the restriction (\ref{eq:reduced1}), 
we will derive the exact gauge transformation laws in \cite{ArItOh2}. 
We will denote the analytic gauge parameter as 
\begin{eqnarray}
\Lambda(\zeta, u) 
&=& 
        \chi (x_A)
+ \bar{\theta}^{+}_{\dot{\alpha}} \lambda^{(0,1)}{}^{\dot{\alpha}} (x_A,u)
        + \theta^{+ \alpha} \lambda^{(1,0)}_{\alpha} (x_A,u)
        + (\bar{\theta}^{+})^2 \lambda^{(0,2)} (x_A, u)
        \nonumber\\
&&{} 
        + (\theta^{+})^2 \lambda^{(2,0)} (x_A, u)
        + \theta^{+}\!\sigma^\mu \bar{\theta}^{+} 
                \lambda_\mu^{(1,1)} (x_A, u)
        + (\bar{\theta}^{+})^2 \theta^{+}{}^{\alpha} 
                \lambda^{(1,2)}_{\alpha} (x_A, u)
        \nonumber\\
&&{} 
        + (\theta^{+})^2 \bar{\theta}^{+}_{\dot{\alpha}} 
                \lambda^{(2,1)}{}^{\dot{\alpha}} (x_A, u)
        + (\theta^{+})^2 (\bar{\theta}^{+})^2 
                \lambda^{(2,2)} (x_A, u) 
, 
\label{eq:AnalyticGaugeParam} 
\end{eqnarray} 
where $\lambda^{(n,m)} (x_A, u)$ 
is the $(\theta^{+})^{n} (\bar{\theta}^{+})^{m}$-component.

The equations to determine the deformed gauge transformation 
are given in \cite{ArItOh2}: 
\begin{eqnarray}
	- 2 i \delta^{*}_{\Lambda} A_\mu 
&=& 
	2i \partial_\mu \chi
	+ 2 \sqrt{2} i C^{(+-)}{}^{\alpha\beta} 
		(\sigma_\mu \bar{\sigma}^\nu \varepsilon )_{\alpha\beta} 
		\partial_\nu \chi \ \bar{\phi}
		\nonumber\\
&&{} 
	- \partial^{++} \lambda_\mu^{(1,1)} 
	- \sqrt{2} C^{++}{}^{\alpha\beta} 
		(\sigma_\mu \bar{\sigma}^\nu \varepsilon )_{\alpha\beta} 
		\lambda_\nu^{(1,1)} \bar{\phi},
	\label{eq:Gauge(1,1)Comp}
	\\
\sqrt{2} i \delta^{*}_{\Lambda} \phi 	
&=&
	- 2 i C^{(+-)}{}^{\alpha\beta} 
		(\sigma_\mu \bar{\sigma}^\nu \varepsilon )_{\alpha\beta} 
		\partial_\nu \chi \ A^\mu 
		\nonumber\\
&&{}
	- C^{++}{}^{\alpha\beta} 
		(\sigma_\mu \bar{\sigma}^\nu \varepsilon )_{\alpha\beta} 
		\lambda_\nu^{(1,1)} A^\mu 
	- \partial^{++} \lambda^{(0,2)}
	, 
	\label{eq:Gauge(0,2)Comp}
	\\
3 \delta^{*}_{\Lambda} D^{ij} u^-_i u^-_j 
&=& 
	4 \sqrt{2} C^{--}{}^{\mu\nu} \partial_\mu \chi \ \partial_\nu \bar{\phi} 
		\nonumber\\ 
&&{} 
	- i \partial^\mu \lambda_\mu^{(1,1)} 
	- \sqrt{2} i C^{(+-)}{}^{\alpha\beta} 
		(\sigma^\mu \bar{\sigma}^\nu \varepsilon )_{\alpha\beta} 
		\partial_\nu (\lambda_\mu^{(1,1)} \bar{\phi} )
	- \partial^{++} \lambda^{(2,2)}
	, 
	\label{eq:Gauge(2,2)Comp}
	\\
4 \delta^{*}_{\Lambda} \psi^{i}_{\alpha} u^-_i 
&=& 
	- 4 C^{(+-)}{}^{\beta\gamma} 
		(\sigma_\mu \bar{\sigma}^\nu \varepsilon )_{\beta\gamma} 
		\partial_\nu \chi \ (\sigma^\mu \bar{\psi}^{i})_\alpha u^-_i 
	- 2 i C^{++}{}^{\beta\gamma} 
		(\sigma_\mu \bar{\sigma}^\nu \varepsilon )_{\beta\gamma} 
		\lambda_\nu^{(1,1)} (\sigma^\mu \bar{\psi}^{i})_\alpha u^-_i 
		\nonumber\\
&&{}
	- \partial^{++} \lambda_\alpha^{(1,2)}
	- 2 \sqrt{2} (\varepsilon C^{++} \lambda^{(1,2)} )_\alpha \bar{\phi}
	. 
\label{eq:Gauge(1,2)Comp}
\end{eqnarray}
Here we have already set $C_s=0$. 
The transformation laws for $\bar{\phi}$ and $\bar{\psi}_i$ are not deformed, 
because their equations are not affected by the $*$-product. 

Now we take into account the restriction (\ref{eq:reduced1}). 
We have 
$C^{(+-)}{}^{\alpha\beta} = C_{11}^{\alpha\beta} u^{+1}u^{-1}$ and so on. 
First we find that eq.(\ref{eq:Gauge(1,1)Comp}) is solved 
in terms of power series of $u^{+1}$ and $u^{-1}$ 
by 
\begin{eqnarray} 
\delta_\Lambda^* 
	A_\mu 
&=& 
	- \partial_\mu \chi 
	, 
	\label{eq:GaugeTr(1,1)} 
	\\ 
\lambda_\mu^{(1,1)} 
&=& 
	\sum_{n=1}^{\infty} \lambda_\mu^{(n)} \ 
		(u^{+1})^{n-1} (u^{-1})^{n+1} 
, 
	\label{eq:GaugeLambda(1,1)} 
\end{eqnarray} 
where 
\begin{equation} 
\lambda_\mu^{(n)} 
\equiv 
	- i { (-2\sqrt{2})^n \over (n+1)! } 
		( 
		\overbrace{ C_{11} \varepsilon C_{11} \cdots \varepsilon C_{11} }^{n} 
		)^{\alpha\beta} 
		(\sigma_\mu \bar{\sigma}^\nu \varepsilon )_{\alpha\beta} 
		\bar{\phi}^n 
		\partial_\nu \chi 
	. 
\end{equation} 
Note that the harmonic expansion of $(u^{+1})^{n} (u^{-1})^{m}$ 
is 
\begin{equation} 
(u^{+1})^{n} (u^{-1})^{m} 
= 
	\overbrace{u^{+(1}\cdots u^{+1}}^{n} 
	\overbrace{u^{-1}\cdots u^{-1)}}^{m} 
= 
	\delta_{(i_1}^1 \cdots \delta_{i_n}^1 \delta_{j_1}^1 \cdots \delta_{j_m)}^1 
		u^{+(i_1}\cdots u^{+ i_n} u^{-j_1}\cdots u^{-j_m)} 
	. 
\end{equation} 
Eq.(\ref{eq:GaugeTr(1,1)}) is directly checked by substituting $\lambda_\mu^{(1,1)}$ 
into the right hand side of eq.(\ref{eq:Gauge(1,1)Comp}),  
noting that $\lambda_\mu^{(n)}$ obeys 
\begin{equation} 
C_{11}^{\alpha\beta} 
	(\sigma_\mu \bar{\sigma}^\nu \varepsilon )_{\alpha\beta} 
	\lambda_\nu^{(n)} \bar{\phi} 
= 
	- {n+2 \over \sqrt{2} } \lambda_\mu^{(n+1)} 
		\qquad (n \ge 1).  
\end{equation} 

Substituting eq.(\ref{eq:GaugeLambda(1,1)}) into (\ref{eq:Gauge(0,2)Comp}) , 
we find 
that all the terms in the right hand side of eq.(\ref{eq:Gauge(0,2)Comp}) 
is gauged away with $\lambda^{(2,0)}$, because 
every term has 
at least one $u^+$. 
This leads to 
\begin{equation} 
\delta_\Lambda^* \phi = 0
. 
\label{eq:GaugeTr(0,2)} 
\end{equation} 
Although the precise form of $\lambda^{(0,2)}$ is not needed, 
it is determined as  
\begin{eqnarray} 
\lambda^{(0,2)} 
&=& 
	- i C_{11}^{\alpha\beta} 
		(\sigma^\mu \bar{\sigma}^\nu \varepsilon )_{\alpha\beta} 
		\partial_\nu \chi A_\mu \ 
		(u^{-1})^{2} 
		\nonumber\\ 
&&{} 
	+ \sum_{n=2}^{\infty} 
		{ -1 \over (n+1) } 
		C_{11}^{\alpha\beta} 
		(\sigma^\mu \bar{\sigma}^\nu \varepsilon )_{\alpha\beta} 
		\lambda_\nu^{(n-1)} A_\mu \ 
		(u^{+1})^{n-1} (u^{-1})^{n+1} 
. 
	\label{eq:GaugeLambda(0,2)} 
\end{eqnarray} 
Similarly, from eq.(\ref{eq:Gauge(2,2)Comp}) we find 
\begin{eqnarray} 
\delta^{*}_{\Lambda} D_{ij} 
&=& 
	\cases{
	2 \sqrt{2} C_{11}^{\mu\nu} \partial_\mu \chi \partial_\nu \bar{\phi}
	,  
	& $(i,j)=(1,1)$ \cr 
	0 
	, 
	& $(i,j)\ne(1,1)$.  
	}
	\label{eq:GaugeTr(2,2)} 
\end{eqnarray} 
This comes from the fact that 
only the first term in the r.h.s. does not contain $u^+$ and 
it contributes to $\delta_{\Lambda}^* D_{11}$  
because it is proportional to $(u^{-1})^2$. 
$\lambda^{(2,2)}$ is then determined as  
\begin{equation} 
\lambda^{(2,2)} 
= 
	- i \sum_{n=2}^{\infty} 
		\partial^\mu \lambda_\mu^{(n)} \ 
		(u^{+1})^{n-2} (u^{-1})^{n+2} 
	. 
	\label{eq:Lambda(2,2)} 
\end{equation}

Eq.(\ref{eq:Gauge(1,2)Comp}) is solved by 
\begin{eqnarray} 
\delta^{*}_{\Lambda} \psi^{i}_{\alpha}  
&=& 
	\delta^i_2 \frac23 (\varepsilon C_{11} \sigma^\mu \bar{\psi}^1 )_{\alpha} 
		\partial_\mu \chi  
	, 
	\\
\lambda^{(1,2)}_\alpha 
&=& 
	\frac 43 
		(\sigma^\mu \bar{\psi}_i )_\alpha C_{11}^{\gamma\delta} 
		(\sigma_\mu \bar{\sigma}^\nu \varepsilon )_{\gamma\delta} 
		\partial_\nu \chi \ 
		u^{-(i}u^{-1}u^{-1)} 
		\nonumber\\ 
&&{} 
	+ \sum_{n=2}^{\infty} 
		i (\sigma^\mu \bar{\psi}_i )_\alpha C_{11}^{\gamma\delta} 
		(\sigma_\mu \bar{\sigma}^\nu \varepsilon )_{\gamma\delta} 
		\lambda_\nu^{(n-1)} 
		\nonumber\\ 
&&{} \qquad \times 
		\biggl[ 
		 a_n 
			\overbrace{u^{+(1}\dots u^{+1}}^{n-1} 
			\overbrace{u^{-i}\dots u^{-1)}}^{n+2}  
		+  b_n 
			\epsilon^{i1} (u^{+1})^{n-2} (u^{-1})^{n+1} 
		\biggr] 
	, 
	\label{eq:GaugeLambda(1,2)} 
\end{eqnarray} 
with $ a_n$ and $ b_n$ being appropriate constants. 
Indeed, substituting (\ref{eq:GaugeLambda(1,1)}) and (\ref{eq:GaugeLambda(1,2)}) 
in appendix \ref{app:UsefulFormulas} 
into eq.(\ref{eq:Gauge(1,2)Comp}) 
and using the formulas (\ref{eq:(I-6)}) and (\ref{eq:(IV-1)}) 
we can see that only the first term in the r.h.s. of (\ref{eq:Gauge(1,2)Comp}) 
gives a contribution to $\delta_\Lambda^* \psi^2$ that cannot be gauged away; 
the other terms are absorbed into $\lambda^{(1,2)}_\alpha$ 
(with the appropriate choice of $ a_n$ and $ b_n$). 

As described above,  
the exact deformed gauge transformation is determined as  
\begin{eqnarray}
\delta^{*}_{\Lambda} A_\mu 
&=& 
	- \partial_\mu \chi 
	, 
	\nonumber\\ 
\delta^{*}_{\Lambda} \psi^{2}_{\alpha}  
&=& 
	- \frac23 (\varepsilon C_{11} \sigma^\mu \bar{\psi}_2 )_{\alpha} 
		\partial_\mu \chi  
	, 
	\nonumber\\ 
\delta^{*}_{\Lambda} D_{11} 
&=& 
	- 2 \sqrt{2} C_{11}^{\mu\nu} \partial_\mu (\partial_\nu \chi \bar{\phi}) 
	, 
	\nonumber\\ 
\delta^{*}_{\Lambda} (\mbox{others})
&=& 
	0
	. 
\label{eq:ReducedGaugeTr1}
\end{eqnarray} 
Note that this result agrees with the deformed gauge transformation 
for the generic $C_{ij}$ in \cite{ArItOh2}, 
after setting $C_{ij}^{\alpha\beta} = 
C_{11}^{\alpha\beta}
 \delta_i^1 \delta_j^1$. 
For example, 
$\delta_\Lambda^* A_\mu$ was determined as  
$ 
\delta^{*}_{\Lambda} A_\mu 
= 
	- \{ 1 + f(\bar{\phi}) \bar{\phi} \} \partial_\mu \chi 
$ 
with $f(\bar{\phi})$ being a function proportional to $C_{12}$ or $C_{21}$ 
(see \cite{ArItOh2} for the definition of $f$), 
so that $f(\bar{\phi})=0$ when 
$C_{ij}=C_{11}\delta^1_i \delta^1_j$,  
which leads to eq.(\ref{eq:GaugeTr(1,1)}). 

The determination of 
the exact ${\cal N}=(1,1/2)$ supersymmetry transformation 
is accomplished in a similar way as the one described above, 
since the deformation of the transformation laws comes from 
the associated gauge transformation to preserve the WZ gauge. 
We will give the result in the next subsection. 
The details of the derivation are found 
in appendix \ref{app:DeterminationOfN=(1,1/2)}.

\subsection{Exact ${\cal N}=(1,\, 1/2)$ supersymmetry transformation} 
\label{sect:ExactN=(1,1/2)} 
Since the theory is expected to have ${\cal N}=(1,\, 1/2)$ supersymmetry, 
we will concentrate on this symmetry in the following. 
For later convenience, we will split 
the ${\cal N}=(1,\, 1/2)$ supersymmetry transformation 
generated by $Q^1$, $Q^2$ and $\bar{Q}_2$ 
into the ${\cal N}=(1,0)$ transformation generated by $Q^1,Q^2$ 
and ${\cal N}=(0,1/2)$ by $\bar{Q}_2$. 
The ${\cal N}=(1,0)$ transformation is defined by 
\begin{equation}
\delta^{*}_{\xi} V^{++}_{WZ} 
= 
        \tilde{\delta}_{\xi} V^{++}_{WZ}
        + \delta^{*}_\Lambda V^{++}_{WZ} 
, 
\label{eq:DeformedSUSY(1,0)}
\end{equation}
and the ${\cal N}=(0,1/2)$ transformation is 
\begin{equation}
\delta^{*}_{\bar{\xi}^2} V^{++}_{WZ} 
= 
        \tilde{\delta}_{\bar{\xi}^2} V^{++}_{WZ} 
        + \delta^{*}_{\Lambda'} V^{++}_{WZ} 
. 
\label{eq:DeformedSUSY(0,1/2)}
\end{equation}
Here 
\begin{equation} 
\tilde{\delta}_{\xi}V^{++}_{WZ}
= 
	\xi_i Q^i V^{++}_{WZ}
= 
	\left(
	- \xi^{+\alpha}Q^{-}_{\alpha}
	+ \xi^{-\alpha}Q^{+}_{\alpha}
	\right) 
	V^{++}_{WZ}
, 
\end{equation} 
\begin{equation} 
\tilde{\delta}_{\bar{\xi}^2}V^{++}_{WZ}
= 
	\bar{\xi}^i \bar{Q}_i V^{++}_{WZ}
= 
	\left(
	\bar{\xi}^{+}_{\dot{\alpha}} \bar{Q}^{-}{}^{\dot{\alpha}}
	- \bar{\xi}^{-}_{\dot{\alpha}} \bar{Q}^{+}{}^{\dot{\alpha}} 
	\right) 
	V^{++}_{WZ}
	\label{eq:tildesusy(0,1/2)}
\end{equation} 
and  $Q^{+}_{\alpha}= {\partial\over\partial\theta^{-\alpha}}
-2i \sigma^{\mu}_{\alpha\dot{\alpha}}\bar{\theta}^{+\dot{\alpha}}
{\partial\over\partial x^{\mu}_{A}}$, 
$Q^{-}_{\alpha}=-{\partial\over\partial \theta^{+\alpha}}$, 
$\bar{Q}^{+}_{\dot{\alpha}}={\partial\over\partial\bar{\theta}^{-\dot{\alpha}}}
+2i \theta^{+\alpha}\sigma^{\mu}_{\alpha\dot{\alpha}}
	{\partial\over \partial x^{\mu}_{A}}$, 
$\bar{Q}^{-}_{\dot{\alpha}}=
-{\partial\over\partial\bar{\theta}^{+\dot{\alpha}}}$. 
Note that $\bar{\xi}^1$ is implicitly set to be zero 
in eq.(\ref{eq:tildesusy(0,1/2)}), 
because we are now considering the ${\cal N}=(1,\, 1/2)$ supersymmetry. 
$\delta^{*}_{\Lambda} V^{++}_{WZ}$ ($\delta^{*}_{\Lambda'} V^{++}_{WZ}$) is 
a deformed gauge transformation of $V^{++}_{WZ}$ 
with an appropriate analytic gauge parameter 
$\Lambda(\zeta,u)$ ($\Lambda'(\zeta,u)$) 
to retain the WZ gauge. 

The exact ${\cal N}=(1,0)$ transformation generated by $Q^1$ and $Q^2$ 
is given below 
(for the derivation, see appendix \ref{app:DeterminationOfN=(1,0)}): 
\begin{eqnarray}
\delta_\xi^* \phi 
&=& 
	- \sqrt{2} i \xi^{i} \psi_{i}
	+ \frac83 i ( \xi_2 \varepsilon C_{11} \psi^1 ) \bar{\phi} 
	- {2 \sqrt{2} \over 3} i 
		( \xi_2 \varepsilon C_{11} \sigma^\nu \bar{\psi}_2 ) 
		A_\nu
	, 
	\nonumber\\
\delta_\xi^* \bar{\phi}
&=& 
	0 
	, 
	\nonumber\\
\delta_\xi^* A_\mu 
&=& 
	i \xi^{i} \sigma_\mu \bar{\psi}_{i}
	+ 2 \sqrt{2}i ( \xi^1 \varepsilon C_{11} \sigma_\mu \bar{\psi}^1 ) 
		\bar{\phi}
	, 
	\nonumber\\
\delta^{*}_{\xi} \psi^{\alpha i} 
&=& 
	- (\xi^{i} \sigma^{\mu\nu} )^{\alpha} F_{\mu\nu}
	- D^{ij} \xi^\alpha_j 
	+ \sqrt{2} C_{11}^{\mu\nu} ( \xi^{(i} \sigma_{\mu\nu} )^\alpha D^{11)} 
		\bar{\phi} 
	- i C_{11}^{\mu\nu} ( \xi^{(i} \sigma_{\mu\nu} )^\alpha 
		(\bar{\psi}^{1} \bar{\psi}^{1)})  
			\nonumber\\
&&{} 
	+ \delta^i_2 \left\{  
		2 \sqrt{2} (\xi^{1} \varepsilon C_{11} 
			\sigma^{\mu\nu} )^\alpha 
		+ {2\sqrt{2}\over 3} (\xi^{1} \sigma^{\mu\nu} \varepsilon 
			C_{11} )^\alpha 
		+ \sqrt{2} C_{11}^{\mu\nu} \xi^{\alpha 1} 
		\right\} 
		\bar{\phi} F_{\mu\nu} 
			\nonumber\\
&&{} 
	- \delta^i_2 \left\{ 
		{4\sqrt{2} \over 3} (\xi^{1} 
			\sigma^{\mu\nu} \varepsilon C_{11} )^\alpha 
		+ 2 \sqrt{2} C_{11}^{\mu\nu} \xi^{\alpha 1} 
		\right\} 
		\partial_\mu \bar{\phi} A_\nu 
	+ \delta^i_2 {2\sqrt{2} \over 3} 
		(\xi^{1} \varepsilon C_{11} )^\alpha 
		\partial^\mu \bar{\phi} A_\mu 
	\nonumber\\ 
&&{} 
	+ \delta^i_2 \det C_{11} 
		\left\{ 
		{8\over 3}
		 \xi^{\alpha 1} D^{11} \bar{\phi}^2 
		- 
		{8\sqrt{2}\over 3}
		 i 
		\xi^{\alpha 1} 
		(\bar{\psi}^1 \bar{\psi}^1) \bar{\phi}
		\right\}  
	, 
	\nonumber\\
\delta_\xi^* \bar{\psi}_{\dot{\alpha} i}  
&=& 
	\sqrt{2} (\xi_{i} \sigma^\nu )_{\dot{\alpha}} 
		\partial_\nu \bar{\phi} 
	+ 2 
		\delta^{1}_{i}
		(\xi_2 \varepsilon C_{11} \sigma^\nu)_{\dot{\alpha}} 
		\partial_\nu (\bar{\phi}^2) 
	, 
	\nonumber\\
\delta_\xi^* D^{11} 
&=& 
	- 2 i \xi^{1} \sigma^\nu \partial_\nu \bar{\psi}^{1}
	, 
	\nonumber\\
\delta_\xi^* D^{12} 
&=& 
	\delta_\xi^* D^{21} 
= 
	- 2 i \xi^{(1} \sigma^\nu \partial_\nu \bar{\psi}^{2)}
	+ 2 \sqrt{2} i 
		\xi^{1} \varepsilon C_{11} \sigma^\nu 
		\partial_\nu (\bar{\phi} \bar{\psi}^{1}) 
	, 
	\nonumber\\
\delta_\xi^* D^{22} 
&=& 
	- 2 i \xi^{2} \sigma^\nu \partial_\nu \bar{\psi}^{2}
	+ 4 \sqrt{2} i 
		\xi^{(1} \varepsilon C_{11} \sigma^\nu 
		\partial_\nu (\bar{\phi} \bar{\psi}^{2)}) 
	+ 8 i \det C_{11} \xi^{1} \sigma^\nu 
		\partial_\nu (\bar{\phi}^2 \bar{\psi}^{1})
	. 
\label{eq:reducedsustr1}
\end{eqnarray} 
Note that we have used 
$(\varepsilon C_{11} \varepsilon C_{11})_\alpha{}^\beta 
= - \delta_\alpha^\beta \det C_{11}$.

The ${\cal N}=(0,1/2)$ supersymmetry transformation 
generated by $\bar{Q}_2$ is as follows 
(these are read from 
eqs.(\ref{eq:N=(0,1):(0,2)})--(\ref{eq:N=(0,1):(2,2)}) 
in appendix \ref{app:DeterminationOfN=(0,1/2)} by setting $\bar{\xi}^1=0$)
: 
\begin{eqnarray} 
&& 
\delta_{\bar{\xi}^2}^* \phi 
= 
	0
	,
	\nonumber\\ 
&& 
\delta_{\bar{\xi}^2}^* \bar{\phi}
= 
	\sqrt{2} i \bar{\xi}^2 \bar{\psi}^1 
	,
	\nonumber\\ 
&& 
\delta_{\bar{\xi}^2}^* A_\mu 
= 
	- i \bar{\xi}^2 \bar{\sigma}_\mu \psi^1 
	, 
	\nonumber\\ 
&& 
\delta_{\bar{\xi}^2}^* \psi^{\alpha i} 
= 
	\delta^i_2 
	\left\{ 
	- \sqrt{2} ( \bar{\xi}^2 \bar{\sigma}^\mu )^\alpha \partial_\mu \phi 
	- \frac23 D^{11} 
		( \bar{\xi}^2 \bar{\sigma}^\mu \varepsilon C_{11} )^\alpha A_\mu 
	+ \frac43 i \bar{\xi}^2 \bar{\psi}^1 ( \psi^1 \varepsilon C_{11} )^\alpha 
	\right\} 
	,
	\nonumber\\ 
&& 
\delta_{\bar{\xi}^2}^* \bar{\psi}_{\dot{\alpha} i} 
= 
	\delta_i^1 ( \bar{\xi}^2 \bar{\sigma}^{\mu\nu} )_{\dot{\alpha}} F_{\mu\nu} 
	- \bar{\xi}_{\dot{\alpha}}^{2} D_{i2} 
	, 
	\nonumber\\ 
&& 
\delta_{\bar{\xi}^2}^* D^{11} 
= 
	0 
	,
	\nonumber\\ 
&& 
\delta_{\bar{\xi}^2}^* D^{12} 
= 
	\delta_{\bar{\xi}^2}^* D^{21} 
= 
	- i \partial_\mu ( \bar{\xi}^{2} \bar{\sigma}^\mu \psi^{1} ) 
	,
	\nonumber\\ 
&& 
\delta_{\bar{\xi}^2}^* D^{22} 
= 
	- 2 i \partial_\mu ( \bar{\xi}^{2} \bar{\sigma}^\mu \psi^{2} ) 
	\nonumber\\ 
&&{} \quad 
	- \frac 23 i \partial_\mu 
		\left[
		2 ( \bar{\xi}^{2} \bar{\sigma}^\nu \varepsilon C_{11} \sigma^\mu 
			\bar{\psi}^{1} ) A_\nu 
		- C_{11}^{\alpha\beta} 
			(\sigma^\nu \bar{\sigma}^\mu \varepsilon)_{\alpha\beta} 
			A_\nu \bar{\xi}^{2} \bar{\psi}^{1} 
		+ \sqrt{2} ( \bar{\xi}^{2} \bar{\sigma}^\mu \varepsilon C_{11} 
			\psi^{1} ) \bar{\phi} 
		\right] 
	. 
\label{eq:N=(0,1/2)}
\end{eqnarray} 

From the exact gauge transformation (\ref{eq:ReducedGaugeTr1}), 
we easily find a field redefinition 
that leads to the canonical gauge transformation: 
\begin{eqnarray} 
\hat{\psi}^2_\alpha 
&\equiv& 
	\psi^2_\alpha 
	- \frac 23 (\varepsilon C_{11} \sigma^\mu \bar{\psi}_2 )_\alpha A_\mu 
	, \\ 
\hat{D}_{11} 
&\equiv& 
	D_{11} 
	- 2 \sqrt{2} C_{11}^{\mu\nu} A_\nu \partial_\mu \bar{\phi} 
	, 
\end{eqnarray}
and the other newly defined fields with hat 
are the same as the original fields.  

Using this field redefinition, 
the exact ${\cal N}=(1,\, 1/2)$ transformation becomes as follows. 
The ${\cal N}=(1,0)$ supersymmetry transformation is 
\begin{eqnarray}
\delta^{*}_{\xi} \hat{\phi} 
&=& 
	- \sqrt{2} i \xi^{i} \hat{\psi}_{i}
	+ \frac83 i ( \xi_2 \varepsilon C_{11} \hat{\psi}^1 ) \hat{\bar{\phi}}  
	, 
	\nonumber\\ 
\delta^{*}_{\xi} \hat{\bar{\phi}}
&=& 
	0 
	, 
	\nonumber\\ 
\delta^{*}_{\xi} \hat{A}_\mu 
&=& 
	i \xi^{i} \sigma_\mu \hat{\bar{\psi}}{}_{i}
	+ 2 \sqrt{2}i ( \xi^1 \varepsilon C_{11} \sigma_\mu \hat{\bar{\psi}}{}^1 ) 
		\hat{\bar{\phi}}
	, 
	\nonumber\\ 
\delta^{*}_{\xi} \hat{\psi}^{\alpha i} 
&=& 
	- (\xi^{i} \sigma^{\mu\nu} )^{\alpha} \hat{F}_{\mu\nu}
	- \hat{D}^{ij} \xi^\alpha_j 
	+ \sqrt{2} C_{11}^{\mu\nu} ( \xi^{(i} \sigma_{\mu\nu} )^\alpha \hat{D}^{11)} 
		\hat{\bar{\phi}} 
	- i C_{11}^{\mu\nu} ( \xi^{i} \sigma_{\mu\nu} )^\alpha 
		(\hat{\bar{\psi}}{}^{1} \hat{\bar{\psi}}{}^{1})  
			\nonumber\\
&&{} 
	+ \delta^i_2 \left\{  
		2 \sqrt{2} (\xi^{1} \varepsilon C_{11} 
			\sigma^{\mu\nu} )^\alpha 
		+ {2\sqrt{2}\over 3} (\xi^{1} \sigma^{\mu\nu} \varepsilon 
			C_{11} )^\alpha 
		+ \sqrt{2} C_{11}^{\mu\nu} \xi^{\alpha 1} 
		\right\} 
		\hat{\bar{\phi}} \hat{F}_{\mu\nu} 
			\nonumber\\
&&{} 
	+ \frac 83 \delta^i_2 \det C_{11} 
		 \xi^{\alpha 1} 
		\left\{ 
		\hat{D}^{11} \hat{\bar{\phi}}{}^2 
		- 2 \sqrt{2} i ( \hat{\bar{\psi}}{}^1 \hat{\bar{\psi}}{}^1 ) 
			\hat{\bar{\phi}}
		\right\} 
	, 
	\nonumber\\ 
\delta^{*}_{\xi} \hat{\bar{\psi}}{}_{\dot{\alpha} i}  
&=& 
	\sqrt{2} (\xi_{i} \sigma^\nu )_{\dot{\alpha}} 
		\partial_\nu \hat{\bar{\phi}} 
	+ 2 
		\delta^{1}_{i}
		(\xi_2 \varepsilon C_{11} \sigma^\nu)_{\dot{\alpha}} 
		\partial_\nu (\hat{\bar{\phi}}{}^2) 
	, 
	\nonumber\\ 
\delta^{*}_{\xi} \hat{D}^{11} 
&=& 
	- 2 i \xi^{1} \sigma^\nu \partial_\nu \hat{\bar{\psi}}{}^{1}
	, 
	\nonumber\\
\delta^{*}_{\xi} \hat{D}^{12} 
&=& 
\delta^{*}_{\xi} \hat{D}^{21} 
= 
	- 2 i \xi^{(1} \sigma^\nu \partial_\nu \hat{\bar{\psi}}{}^{2)}
	+ 2 \sqrt{2} i 
		\xi^{1} \varepsilon C_{11} \sigma^\nu 
		\partial_\nu (\hat{\bar{\phi}} \hat{\bar{\psi}}{}^{1}) 
	, 
	\nonumber\\
\delta^{*}_{\xi} \hat{D}^{22} 
&=& 
	- 2 i \xi^{2} \sigma^\nu \partial_\nu \hat{\bar{\psi}}{}^{2}
	+ 4 \sqrt{2} i 
		\xi^{(1} \varepsilon C_{11} \sigma^\nu 
		\partial_\nu (\hat{\bar{\phi}} \hat{\bar{\psi}}{}^{2)}) 
	- 2 \sqrt{2} i ( \xi^i \varepsilon C_{11} \sigma^\nu \hat{\bar{\psi}}{}_i ) 
		\partial_\nu \hat{\bar{\phi}} 
		\nonumber\\ 
&&{} 
	+ 8 i \det C_{11} 
		\left\{ 
		\xi^{1} \sigma^\nu 
			\partial_\nu (\hat{\bar{\phi}}{}^2 \hat{\bar{\psi}}{}^{1}) 
		+ ( \xi^{1} \sigma^\nu \hat{\bar{\psi}}{}^{1} ) 
			\partial_\nu \hat{\bar{\phi}} \hat{\bar{\phi}}
		\right\} 
	, 
\end{eqnarray} 
where 
$
\hat{F}_{\mu\nu} 
\equiv 
	\partial_\mu \hat{A}_\nu - \partial_\nu \hat{A}_\mu
$. 
The ${\cal N}=(0,\ 1/2)$ supersymmetry transformation becomes 
\begin{eqnarray} 
\delta_{\bar{\xi}^2}^* \hat{\phi} 
&=& 
	0
	,
	\nonumber\\ 
\delta_{\bar{\xi}^2}^* \hat{\bar{\phi}}
&=& 
	\sqrt{2} i \bar{\xi}^2 \hat{\bar{\psi}}{}^1 
	,
	\nonumber\\ 
\delta_{\bar{\xi}^2}^* \hat{A}_\mu 
&=& 
	- i \bar{\xi}^2 \bar{\sigma}_\mu \hat{\psi}^1 
	, 
	\nonumber\\ 
\delta_{\bar{\xi}^2}^* \hat{\psi}^{\alpha i} 
&=& 
	\delta^i_2 
	\left\{ 
	- \sqrt{2} ( \bar{\xi}^2 \bar{\sigma}^\mu )^\alpha \partial_\mu \hat{\phi} 
	+ \frac83 i \bar{\xi}^2 \hat{\bar{\psi}}{}^1 
		( \hat{\psi}^1 \varepsilon C_{11} )^\alpha 
	\right\} 
	,
	\nonumber\\ 
\delta_{\bar{\xi}^2}^* \hat{\bar{\psi}}{}_{\dot{\alpha} i} 
&=& 
	\delta_i^1 ( \bar{\xi}^2 \bar{\sigma}^{\mu\nu} )_{\dot{\alpha}} 
		\hat{F}_{\mu\nu} 
	- \bar{\xi}_{\dot{\alpha}}^{2} \hat{D}_{i2} 
	, 
	\nonumber\\ 
\delta_{\bar{\xi}^2}^* \hat{D}^{11} 
&=& 
	0 
	,
	\nonumber\\ 
\delta_{\bar{\xi}^2}^* \hat{D}^{12} 
&=& 
	\delta_{\bar{\xi}^2}^* \hat{D}^{21} 
= 
	- i \partial_\mu ( \bar{\xi}^{2} \bar{\sigma}^\mu \hat{\psi}^{1} ) 
	,
	\nonumber\\ 
\delta_{\bar{\xi}^2}^* \hat{D}^{22} 
&=& 
	- 2 i \partial_\mu ( \bar{\xi}^{2} \bar{\sigma}^\mu \hat{\psi}^{2} ) 
	+ 2 i C_{11}^{\mu\nu} \bar{\xi}^2 \hat{\bar{\psi}}{}^1 \hat{F}_{\mu\nu} 
	\nonumber\\ 
&&{} 
	- 2 \sqrt{2} i ( \bar{\xi}^{2} \bar{\sigma}^\mu \varepsilon C_{11} 
			\hat{\psi}^{1} ) \partial_\mu \hat{\bar{\phi}} 
	- {2 \sqrt{2}\over 3} i \partial_\mu 
		\left[
		( \bar{\xi}^{2} \bar{\sigma}^\mu \varepsilon C_{11} 
			\hat{\psi}^{1} ) \hat{\bar{\phi}} 
		\right] 
	. 
\label{eq:n012transf}
\end{eqnarray} 
In terms of the newly defined fields, 
the action in the WZ gauge should now be invariant 
under the ordinary gauge transformation 
and the above ${\cal N}=(1,\ 1/2)$ supersymmetry transformation. 

Note that 
the $O(C)$ action is exactly 
gauge and ${\cal N}=(1,\ 1/2)$ supersymmetry invariant by itself 
(in order to show this 
we have used $(\bar{\psi}^1)^3=0$ as a result of the $U(1)$ gauge group). 
An immediate consequence of this fact is that  
the terms arising from $(V^{++})^n$ ($n\ge 4$) in the action, 
if there exist, 
have to be also gauge and ${\cal N}=(1,\ 1/2)$ supersymmetry invariant 
by themselves. 

Using the field redefinitions (\ref{eq:identification2}) and 
(\ref{eq:identification3}) and transformation (\ref{eq:n012transf}), 
we find that the action
(\ref{eq:n12action1}) is invariant under the ${\cal N}=(0,1/2)$
transformation:
\begin{eqnarray} 
&& 
\delta_{\bar{\eta}}^* A 
= 
	0 
	, 
	\quad 
\delta_{\bar{\eta}}^* \bar{A} 
= 
	\sqrt{2} \bar{\eta} \bar{\lambda} 
	, 
	\nonumber\\ 
&& 
\delta_{\bar{\eta}}^* \psi^\alpha 
= 
	0 
	, 
	\quad 
\delta_{\bar{\eta}}^* \bar{\psi}_{\dot{\alpha}} 
= 
	( \bar{\eta} \bar{\sigma}^{\mu\nu} )_{\dot{\alpha}} v_{\mu\nu} 
	- i \bar{\eta}_{\dot{\alpha}} D 
	, 
	\nonumber\\ 
&& 
\delta_{\bar{\eta}}^* F 
= 
	0
	, 
	\quad 
\delta_{\bar{\eta}}^* \bar{F} 
= 
	\sqrt{2} i \bar{\eta} \bar{\sigma}^\mu \partial_\mu 
	\lambda
	+ 2 
		\bar{\eta} \bar{\sigma}^\mu \varepsilon C \partial_\mu (\psi \bar{A}) 
	, 
	\nonumber\\ 
&& 
\delta_{\bar{\eta}}^* v_\mu 
= 
	- i \bar{\eta} \bar{\sigma}_\mu \psi 
	, 
	\nonumber\\	
&& 
\delta_{\bar{\eta}}^* \lambda^\alpha 
= 
	-
	i \sqrt{2} ( \bar{\eta} \bar{\sigma}^\mu )^\alpha \partial_\mu A 
	- 2 i 
	\bar{\eta} \bar{\lambda} ( \psi \varepsilon C )^\alpha 
	, 
	\quad 
\delta_{\bar{\eta}}^* \bar{\lambda}_{\dot{\alpha}} 
= 
	\sqrt{2} \bar{\eta}_{\dot{\alpha}} F
	, 
	\nonumber\\ 
&& 
\delta_{\bar{\eta}}^* D 
= 
	\bar{\eta} \bar{\sigma}^\mu \partial_\mu \psi  
	. 
\label{eq:etabarsusy1}
\end{eqnarray} 
Thus we have obtained deformed ${\cal N}=(1,\, 1/2)$ supersymmetry 
transformations (\ref{eq:n12susytr1}), (\ref{eq:etasusy1}) and 
(\ref{eq:etabarsusy1}) in
${\cal N}=2$ supersymmetric $U(1)$ gauge theory in the deformed ${\cal N}=1$ superspace.

\section{Conclusions and Discussion}
In this paper, we studied ${\cal N}=2$  supersymmetric $U(1)$ gauge
theory in non(anti)commutative ${\cal N}=1$ and ${\cal N}=2$ harmonic 
superspaces.
We considered the reduction of the deformation parameters in
non(anti)commutative harmonic superspace to the deformed ${\cal N}=1$ 
superspace.
We found that the $O(C)$ action in harmonic superspace reduces to the one
in the deformed ${\cal N}=1$ superspace by the field redefinition.
We calculated the gauge and ${\cal N}=(1,\, 1/2)$
supersymmetry transformations exactly.
Using the field redefinition, we confirmed ${\cal N}=(1,\, 1/2)$
supersymmetry of the action in the deformed ${\cal N}=1$ superspace.

It is known that even in the $U(1)$ gauge group there exists higher
order $C$-corrections to the action 
in the deformed harmonic superspace\cite{ArItOh2,ArIt3}.
It is not clear whether these higher order contributions in the
action disappear after the reduction 
of the deformation parameters(\ref{eq:reduced1}).
Even if these terms exist after the reduction,
these must be gauge- and ${\cal N}=(1,\,1/2)$ supersymmetry invariant by
themselves and reduce to the deformed action 
in the ${\cal N}=1$ superspace formalism by appropriate field
redefinition.
Some detailed analysis will be studied elsewhere. 

In this paper, 
the component ${\cal N}=(1,\,1/2)$ supersymmetry transformation 
for the action in the deformed ${\cal N}=1$ superspace 
has not fixed completely, because it is unclear  
whether the field identification (\ref{eq:identification2}) is exact or not. 
To fix the component transformation, 
it is useful to work with the ${\cal N}=2$ rigid superspace formalism. 
The deformed action could be written in terms of 
the ${\cal N}=2$ chiral superfield 
that describes the ${\cal N}=2$ vector multiplet  
and give directly 
the action in the deformed ${\cal N}=1$ superspace. 
If we can construct 
the appropriately deformed ${\cal N}=2$ chiral superfield,  
the component transformation 
that is not fixed in this paper will be determined.  
This formalism would be also useful to study 
a generalization to non-abelian gauge group. 

Another interesting issue would be 
a central extension of ${\cal N}=(1,\, 1/2)$ 
supersymmetry algebra.
It is also interesting to study properties of 
deformed properties of solitons in such as monopoles and instantons
in such theories \cite{Inst}.

{\bf Acknowledgments}:
T.~A. is supported by 
the Grant-in-Aid for Scientific Research in Priority Areas (No.14046201) 
from the Ministry of Education, Culture, Sports, Science 
and Technology. 
A.~O. is supported by a 21st Century COE Program at 
Tokyo Tech "Nanometer-Scale Quantum Physics" by the 
Ministry of Education, Culture, Sports, Science and Technology.

\appendix

\section{Useful formulas} 
\label{app:UsefulFormulas}
In this appendix, we describe some useful reduction formulas 
of harmonic variables, which we have used 
in sect.\ \ref{sect:N=(1,1/2)SUSY} to determine 
the exact transformation laws in the harmonic superspace formalism. 
They are 
\begin{eqnarray} 
&&
u^{+(1}u^{+1)}
	\overbrace{
	u^{+}{}^{(1} \cdots u^{+}{}^{1} 
	}^{n}
	\overbrace{
	u^{-}{}^{i} u^{-}{}^{j} u^{-}{}^{k} u^{-}{}^{1} \cdots u^{-}{}^{1)} 
	}^{m}
	\nonumber\\
&&\quad {}= 
	\overbrace{
	u^{+}{}^{(1} \cdots u^{+}{}^{1} 
	}^{n+2}
	\overbrace{
	u^{-}{}^{i} u^{-}{}^{j} u^{-}{}^{k} u^{-}{}^{1} \cdots u^{-}{}^{1)}
	}^{m}
	\nonumber\\
&&\quad
	+ { 2m \over (n+m+2)(n+m) } 
		\nonumber\\
&&\qquad \times 
		\Bigl( 
		\epsilon^{i 1} 
			\overbrace{ 
			u^{+}{}^{(1} \cdots u^{+}{}^{1} 
			}^{n+1}
			\overbrace{
			u^{-}{}^{j} u^{-}{}^{k} u^{-}{}^{1} \cdots u^{-}{}^{1)}
			}^{m-1}
		+ (\mbox{cyclic permut. of $(i,j,k)$})
		\Bigr) 
		\nonumber\\
&&\quad 
	+ {2m(m-1)\over (n+m+1)(n+m)^2(n+m-1)} 
		\nonumber\\
&&\qquad \times 
	\Bigl( 
		\epsilon^{i 1} \epsilon^{j 1} 
			\overbrace{ 
			u^{+}{}^{(1} \cdots u^{+}{}^{1} 
			}^{n}
			\overbrace{
			u^{-}{}^{k} u^{-}{}^{1} \cdots u^{-}{}^{1)}
			}^{m-2}
		+ (\mbox{cyclic permut. of $(i,j,k)$})
	\Bigr)
	, 
	\label{eq:(I-6)} 
	\\ 
&&
u^{-k} 
	\overbrace{
	u^{+}{}^{(1} \cdots u^{+}{}^{1} 
	}^{n}
	\overbrace{
	u^{-i} u^{-j} u^{-}{}^{1} \cdots u^{-}{}^{1)} 
	}^{m}
	\nonumber\\
&&\quad {}= 
	\overbrace{
	u^{+}{}^{(1} \cdots u^{+}{}^{1} 
	}^{n}
	\overbrace{
	u^{-i} u^{-j} u^{-k} u^{-}{}^{1} \cdots u^{-}{}^{1)}
	}^{m+1}
	\nonumber\\
&&\quad
	+ { 2n \over (n+m+1)(n+m) } 
		\frac12 \Bigl\{ 
		\epsilon^{ki} 
		\overbrace{ 
		u^{+}{}^{(1} \cdots u^{+}{}^{1} 
		}^{n-1}
		\overbrace{
		u^{-j} u^{-}{}^{1} \cdots u^{-}{}^{1)}
		}^{m}
		+ (i \leftrightarrow j) 
		\Bigr\} 
		\nonumber\\
&&\quad 
	+ {n(n+m-2) \over (n+m+1)(n+m)} 
		\epsilon^{k1} 
		\overbrace{
		u^{+}{}^{(1} \cdots u^{+}{}^{1} 
		}^{n-1}
		\overbrace{
		u^{-i} u^{-j} u^{-}{}^{1} \cdots u^{-}{}^{1)}
		}^{m}
	, 
	\label{eq:(IV-1)} 
	\\ 
&&
u^{-(k}u^{-l)}
	\overbrace{
	u^{+}{}^{(i} u^{+}{}^{1} \cdots u^{+}{}^{1} 
	}^{n}
	\overbrace{
	u^{-}{}^{1} \cdots u^{-}{}^{1)} 
	}^{m}
	\nonumber\\
&&\quad {}= 
	\overbrace{
	u^{+}{}^{(1} \cdots u^{+}{}^{1} 
	}^{n}
	\overbrace{
	u^{-}{}^{i} u^{-}{}^{k} u^{-}{}^{l} u^{-}{}^{1} \cdots u^{-}{}^{1)}
	}^{m+2}
	\nonumber\\
&&\quad
	+ { 2n(n+m-1) \over (n+m+2)(n+m) } 
		\frac12 
		\Bigl\{
		\epsilon^{k 1} 
			\overbrace{ 
			u^{+}{}^{(1} \cdots u^{+}{}^{1} 
			}^{n-1}
			\overbrace{
			u^{-}{}^{l} u^{-}{}^{i} u^{-}{}^{1} \cdots u^{-}{}^{1)}
			}^{m+1}
		+ (k\leftrightarrow l) 
		\Bigr\}
	\nonumber\\
&&\quad
	+ { 2n \over (n+m+2)(n+m) } 
		\frac12 
		\Bigl\{
		\epsilon^{k i} 
			\overbrace{ 
			u^{+}{}^{(1} \cdots u^{+}{}^{1} 
			}^{n-1}
			\overbrace{
			u^{-}{}^{l} u^{-}{}^{1} \cdots u^{-}{}^{1)}
			}^{m+1}
		+ (k\leftrightarrow l) 
		\Bigr\}
	\nonumber\\
&&\quad
	+ { n(n-1)(n+m-2) \over (n+m+1)(n+m)^2 } 
		\epsilon^{l1} \epsilon^{k1} 
			\overbrace{ 
			u^{+}{}^{(1} \cdots u^{+}{}^{1} 
			}^{n-2}
			\overbrace{
			u^{-}{}^{i} u^{-}{}^{1} \cdots u^{-}{}^{1)}
			}^{m}
	\nonumber\\
&&\quad
	+ { 2n(n-1) \over (n+m+1)(n+m)^2 } 
		\frac12 
		\Bigl\{
		\epsilon^{l1} \epsilon^{k i} 
			\overbrace{ 
			u^{+}{}^{(1} \cdots u^{+}{}^{1} 
			}^{n-2}
			\overbrace{
			u^{-}{}^{1} \cdots u^{-}{}^{1)}
			}^{m}
		+ (k\leftrightarrow l) 
		\Bigr\}
	, 
	\label{eq:(III-1)} 
	\\ 
&&
u^{+(1}u^{-1)}
	\overbrace{
	u^{+}{}^{(i} u^{+}{}^{1} \dots u^{+}{}^{1} 
	}^{n}
	\overbrace{
	u^{-}{}^{k} u^{-}{}^{1} \dots u^{-}{}^{1)} 
	}^{m} 
	\nonumber\\
&&\quad {}= 
	\overbrace{
	u^{+}{}^{(i} u^{+}{}^{1} \dots u^{+}{}^{1} 
	}^{n+1}
	\overbrace{
	u^{-}{}^{k} u^{-}{}^{1} \dots u^{-}{}^{1)}
	}^{m+1}
	\nonumber\\
&&\quad
	- { 2(n-m) \over (n+m+2)(n+m) }
		\frac12 
		\Bigl( 
		\epsilon^{i 1} 
		\overbrace{ 
		u^{+}{}^{(1} \dots u^{+}{}^{1} 
		}^{n}
		\overbrace{
		u^{-}{}^{k} u^{-}{}^{1} \dots u^{-}{}^{1)}
		}^{m}
		+ (i \leftrightarrow k)
		\Bigr) 
		\nonumber\\
&&\quad 
	- { 2nm \over (n+m+1)(n+m)^2(n+m-1)} 
		\epsilon^{i 1} \epsilon^{k 1} 
		\overbrace{ 
		u^{+}{}^{(1} \dots u^{+}{}^{1} 
		}^{n-1}
		\overbrace{
		u^{-}{}^{1} \dots u^{-}{}^{1)}
		}^{m-1}
	. 
	\label{eq:(II-5)} 
\end{eqnarray}

\section{${\cal N}=(1,\, 1/2)$ Supersymmetry}
\label{app:DeterminationOfN=(1,1/2)}
Here we will present the details of calculation of the 
exact ${\cal N}=(1,\, 1/2)$ supersymmetry transformation 
of the component fields 
when the deformation parameters are restricted as (\ref{eq:reduced1}). 
The result has been summarized in sect.\ \ref{sect:ExactN=(1,1/2)}. 

\subsection{${\cal N}=(1,0)$ supersymmetry} 
\label{app:DeterminationOfN=(1,0)}

The equations to determine the deformed ${\cal N}=(1,0)$ 
supersymmetry transformation are 
eqs.(\ref{eq:EqsToDetermineDeformedSUSY(0,1)})--%
(\ref{eq:EqsToDetermineDeformedSUSY(2,2)}) which are 
obtained from eq.(\ref{eq:DeformedSUSY(1,0)}) and (\ref{eq:AnalyticGaugeParam}): 
\begin{eqnarray}
0 
&=& 
	2 i (\xi^{+} \sigma^\mu)_{\dot{\beta}} 
			\varepsilon^{\dot{\beta}\dot{\alpha}} A_\mu 
	- \partial^{++} \lambda^{(0,1)}{}^{\dot{\alpha}} 
	- 2 \lambda^{(1,0)}{}^{\alpha} 
			(\varepsilon C^{++} \sigma^\mu )_{\alpha{\dot{\beta}}}
			\varepsilon^{{\dot{\beta}}{\dot{\alpha}}} 
			A_\mu 
	, 
	\label{eq:EqsToDetermineDeformedSUSY(0,1)}
	\\
0 
&=& 
	- 2 \sqrt{2} i \xi^{+}_{\alpha} \bar{\phi} 
	- \partial^{++} \lambda^{(1,0)}_\alpha 
	- 2 \sqrt{2} (\varepsilon C^{++} \lambda^{(1,0)} )_\alpha \bar{\phi} 
	, 
	\label{eq:EqsToDetermineDeformedSUSY(1,0)}
	\\
\sqrt{2} i 
\delta^{*}_{\xi} \phi 
&=& 
	4 \xi^{+} \psi^i u_i^{-}
	+ 4 i \lambda^{(1,0)}{}^{\alpha} (\varepsilon C^{++} \psi^i )_\alpha u_i^{-} 
	\nonumber\\
&&{} 
	- \partial^{++} \lambda^{(0,2)} 
	- C^{++}{}^{\alpha\beta} 
			(\sigma^\nu \bar{\sigma}^\mu \varepsilon )_{\alpha\beta} 
			\lambda_\mu^{(1,1)} A_\nu 
	, 
	\label{eq:EqsToDetermineDeformedSUSY(0,2)}
	\\
- \sqrt{2} i 
\delta^{*}_{\xi} \bar{\phi}
&=& 
	0 
	, 
	\label{eq:EqsToDetermineDeformedSUSY(2,0)}
	\\
- 2 i 
\delta^{*}_{\xi} A_\mu 
&=& 
	4 \xi^{+} \sigma_\mu \bar{\psi}^i u_i^{-}
	+ 4 i \lambda^{(1,0)}{}^{\alpha} 
			( \varepsilon C^{++} \sigma_{\mu} 
			\bar{\psi}^i )_\alpha u_i^{-} 
	\nonumber\\
&&{} 
	- \partial^{++} \lambda_\mu^{(1,1)} 
	- \sqrt{2} C^{++}{}^{\alpha\beta} 
			( \sigma_\mu \bar{\sigma}^\nu \varepsilon)_{\alpha\beta}
			\lambda_\nu^{(1,1)} \bar{\phi}
	, 
	\label{eq:EqsToDetermineDeformedSUSY(1,1)}
	\\
4 
\delta^{*}_{\xi} \psi_{\alpha}^i u_i^{-} 
&=& 
	- 2 (\sigma^\mu \bar{\sigma}^\nu \xi^{-})_{\alpha} 
			\partial_\nu A_\mu 
	+ 6 \xi^{+}_{\alpha} D^{ij} u_i^{-} u_j^{-} 
			\nonumber\\
&&{} 
	- i (\sigma^\nu \partial_\nu \lambda^{(0,1)})_\alpha 
	- 2 \sqrt{2} i ( \varepsilon C^{+-} 
			\sigma^\nu \partial_\nu \lambda^{(0,1)} )_\alpha \bar{\phi} 
			\nonumber\\
&&{} 
	- 6 i (\varepsilon C^{++} \lambda^{(1,0)})_{\alpha} D^{ij} u_i^{-} u_j^{-} 
	+ 2 i \lambda^{(1,0)}{}^{\beta} 
			(\varepsilon C^{+-} \sigma^\nu 
			\bar{\sigma}^\mu \varepsilon)_{\beta\alpha} 
			\partial_\nu A_\mu 
			\nonumber\\
&&{} 
	+ 2 i \partial_\nu \lambda^{(1,0)}_{\alpha} C^{+-}{}^{\beta\gamma} 
			(\sigma^\mu \bar{\sigma}^\nu \varepsilon)_{\beta\gamma} 
			A_\mu 
	- 2 i (\sigma^\mu \bar{\psi}^i )_{\alpha} u_i^{-} 
			C^{++}{}^{\gamma\delta} 
			(\sigma_\mu \bar{\sigma}^\nu \varepsilon)_{\gamma\delta}
			\lambda_\nu^{(1,1)} 
			\nonumber\\
&&{} 
	- \partial^{++} \lambda^{(1,2)}_\alpha 
	- 2 \sqrt{2} ( \varepsilon C^{++} \lambda^{(1,2)} )_\alpha \bar{\phi} 
	- 2 
		(\varepsilon C^{++} \sigma^\mu \lambda^{(2,1)} )_\alpha A_\mu 
	, 
	\label{eq:EqsToDetermineDeformedSUSY(1,2)}
	\\
-4 
\delta^{*}_{\xi} \bar{\psi}^{\dot{\alpha} i} u_i^{-}
&=& 
	2 \sqrt{2} (\xi^{-} \sigma^\mu)_{\dot{\beta}} 
			\varepsilon^{\dot{\beta}\dot{\alpha}} 
			\partial_\mu \bar{\phi} 
	+ i \partial_\mu \lambda^{(1,0)}{}^\alpha 
			\sigma^\mu_{\alpha{\dot{\beta}}} 
			\varepsilon^{\dot{\beta}\dot{\alpha}} 
			\nonumber\\
&&{} 
	+ 2 \sqrt{2} i \partial_\nu 
		\left\{ 
		\lambda^{(1,0)}{}^{\alpha} 
			(\varepsilon C^{+-} \sigma^\nu )_{\alpha{\dot{\beta}}} 
			\varepsilon^{{\dot{\beta}}{\dot{\alpha}}} 
			\bar{\phi} 
		\right\} 
	- \partial^{++} \lambda^{(2,1)}{}^{\dot{\alpha}} 
	, 
	\label{eq:EqsToDetermineDeformedSUSY(2,1)}
	\\
3
\delta^{*}_{\xi} D^{ij} u_i^{-} u_j^{-} 
&=& 
	- 4 i \xi^{-} \sigma^\mu \partial_\mu \bar{\psi}^i u_i^{-}
	+ 4 \partial_\nu 
		\left\{ 
		\lambda^{(1,0)}{}^{\alpha} 
			(\varepsilon C^{+-} \sigma^\nu \bar{\psi}^i )_{\alpha} u_i^{-} 
		\right\} 
			\nonumber\\
&&{} 
	- i \partial^\mu \lambda_\mu^{(1,1)} 
	- \sqrt{2} i C^{+-}{}^{\alpha\beta} 
			( \sigma^\mu \bar{\sigma}^\nu \varepsilon )_{\alpha\beta} 
			\partial_\nu ( \lambda_\mu^{(1,1)} \bar{\phi} )
	- \partial^{++} \lambda^{(2,2)}
. 
\label{eq:EqsToDetermineDeformedSUSY(2,2)}
\end{eqnarray} 
Note that under the restriction (\ref{eq:reduced1}), we have 
$C^{+-}{}^{\alpha\beta} = C_{11}^{\alpha\beta} u^{+1}u^{-1}$ and so on. 

Eq.(\ref{eq:EqsToDetermineDeformedSUSY(1,0)}) is solved by 
\begin{eqnarray} 
\lambda^{(1,0)}{}^\alpha 
&=& 
	2 \sqrt{2} i \xi_i^\alpha \bar{\phi} \ u^{-i} 
	+ \sum_{n=1}^{\infty} 
		2 \sqrt{2} i \xi_i^{(n)}{}^\alpha \bar{\phi} 
		\nonumber\\ 
&&{} \qquad \times 
		\biggl[
		\alpha^{(1,0)}_n 
			\overbrace{u^{+(1}\cdots u^{+1}}^{n} 
			\overbrace{u^{-i}u^{-1}\cdots u^{-1)}}^{n+1} 
		+ \beta^{(1,0)}_n \epsilon^{i1} (u^{+1})^{n-1} (u^{-1})^n
		\biggr]
, 
\label{eq:SUSYLambda(1,0)}
\end{eqnarray} 
where 
\begin{equation} 
\xi_i^{(n)}{}^\alpha  
\equiv 
	( \xi_i \overbrace{\varepsilon C_{11} \cdots \varepsilon C_{11}}^{n}
		)^\alpha 
		( 2 \sqrt{2} \bar{\phi} )^{n} 
\end{equation} 
and 
\begin{eqnarray} 
&&
(\alpha^{(1,0)}_1, \beta^{(1,0)}_1 ) 
= 
	\left( \matrix{ \frac 12, & \frac 23 } \right)
	, 
	\nonumber\\ 
&& 
\alpha^{(1,0)}_n 
= 
	{1\over n+1} \alpha^{(1,0)}_{n-1} 
	, 
	\quad 
\beta^{(1,0)}_n 
= 
	\frac 1n 
		\left(
		{2n \over (2n+1)(2n-1) } \alpha^{(1,0)}_{n-1} 
		+ \beta^{(1,0)}_{n-1} 
		\right) 
. 
\label{eq:NumSeries(1,0)}
\end{eqnarray} 
We then obtain 
$ 
\alpha^{(1,0)}_n 
= 
	{1\over (n+1)!} 
$,  
$\beta^{(1,0)}_n 
= 
	{2\over (n-1)! (2n+1)} 
$.  
To check (\ref{eq:NumSeries(1,0)}), 
we need (\ref{eq:(I-6)}), (\ref{eq:(IV-1)}) 
 and 
\begin{equation} 
2 \sqrt{2} ( \xi_i^{(n)} \varepsilon C_{11} )^\alpha \bar{\phi}  
= 
	\xi_i^{(n+1)}{}^{\alpha} 
. 
\end{equation}

{}From eq.(\ref{eq:EqsToDetermineDeformedSUSY(0,1)}) we find 
\begin{eqnarray} 
\lambda^{(0,1)}_{\dot{\alpha}} 
&=& 
	2 i (\xi_i \sigma^\mu )_{\dot{\alpha}} A_\mu \ u^{-i} 
	\nonumber\\ 
&&{}
	+ \sum_{n=1}^{\infty} 
		2 i ( \xi_i^{(n)} \sigma^\mu )_{\dot{\alpha}} A_\mu 
		\biggl[ 
		\alpha^{(1,0)}_{n} 
				\overbrace{u^{+(1}\cdots u^{+1}}^{n} 
				\overbrace{u^{-i}u^{-1}\cdots u^{-1)}}^{n+1} 
		+ \beta^{(1,0)}_n \epsilon^{i1} 
			(u^{+1})^{n-1} (u^{-1})^n
		\biggr] 
	. 
	\nonumber\\ 
&& 
\label{eq:SUSYLambda(0,1)}
\end{eqnarray}

In eq.(\ref{eq:EqsToDetermineDeformedSUSY(1,1)}), 
$\lambda^{(1,1)}_\mu$ is given by 
\begin{eqnarray} 
\lambda^{(1,1)}_\mu 
&=& 
	2 \xi_i \sigma_\mu \bar{\psi}_j \ u^{-(i}u^{-j)} 
		\nonumber\\ 
&&{} 
	+ \xi_i^{(1)} \sigma_\mu \bar{\psi}_j 
		\left[ 
		2 u^{+(1}u^{-i}u^{-j}u^{-1)}
		+ \frac32  
			\Bigl\{ 
			\epsilon ^{i1} 
				u^{-(j}u^{-1)} 
			+ ( i \leftrightarrow j ) 
			\ \Bigr\}
		\right] 
		\nonumber\\ 
&&{} 
	+ \sum_{n=2}^{\infty} 
		\xi_i^{(n)} \sigma_\mu \bar{\psi}_j 
		\biggl[ 
		\alpha^{(1,1)}_n 
			\overbrace{u^{+(1}\cdots u^{+1}}^{n} 
			\overbrace{u^{-i}u^{-j}u^{-1}\cdots u^{-1)}}^{n+2} 
		\nonumber\\ 
&&{} \qquad 
		+ \beta^{(1,1)}_n 
			\frac12 
			\Bigl\{ 
			\epsilon ^{i1} 
				\overbrace{u^{+(1}\cdots u^{+1}}^{n-1} 
				\overbrace{u^{-j}u^{-1}\cdots u^{-1)}}^{n+1} 
			+ ( i \leftrightarrow j ) 
			\ \Bigr\}
		\nonumber\\ 
&&{} \qquad 
		+ \gamma^{(1,1)}_n 
			\epsilon ^{i1} \epsilon ^{j1} 
			(u^{+1})^{n-2} (u^{-1})^n
		+ \delta^{(1,1)}_n 
			\epsilon^{ji} 
			(u^{+1})^{n-1} (u^{-1})^{n+1} 
		\biggr] 
, 
\label{eq:SUSYLambda(1,1)} 
\end{eqnarray} 
where 
$(\alpha^{(1,1)}_2, \beta^{(1,1)}_2, \gamma^{(1,1)}_2, \delta^{(1,1)}_2)
= ( 1, \frac 83, \frac 85, - \frac 13 )$ and 
\begin{eqnarray} 
&& 
\alpha^{(1,1)}_{n} 
= 
	{1\over n+1} \left( 4 \alpha^{(1,0)}_{n-1} + \alpha^{(1,1)}_{n-1} \right) 
	, \nonumber\\ 
&& 
\beta^{(1,1)}_{n} 
= 
	{1\over n+1} 
		\left( 
		{4n\over 2n-1} \alpha^{(1,0)}_{n-1} 
		+ 4 \beta^{(1,0)}_{n-1} 
		+ {1\over n} \alpha^{(1,1)}_{n-1} 
		+ \beta^{(1,1)}_{n-1} 
		\right) 
	, \nonumber\\ 
&& 
\gamma^{(1,1)}_{n} 
= 
	{1\over n} 
		\left( 
		{4n\over (2n+1)(2n-1)} \alpha^{(1,0)}_{n-1} 
		+ 2 \beta^{(1,0)}_{n-1} 
		\right. \nonumber\\ 
&&\left. {} \qquad \qquad 
		+ {n+1 \over (2n+1)2n(2n-1)} \alpha^{(1,1)}_{n-1} 
		+ {1\over 2n-2} \beta^{(1,1)}_{n-1} 
		+ \gamma^{(1,1)}_{n-1} 
		\right) 
	, \nonumber\\ 
&& 
\delta^{(1,1)}_{n} 
= 
	{1\over n+1} 
		\left( 
		{2(n-1)\over 2n-1} \alpha^{(1,0)}_{n-1} 
		- 2 \beta^{(1,0)}_{n-1} 
		+ \delta^{(1,1)}_{n-1} 
		\right) 
	. 
\end{eqnarray} 
Substituting eq.(\ref{eq:SUSYLambda(1,1)}) to the r.h.s. of 
eq.(\ref{eq:EqsToDetermineDeformedSUSY(1,1)}) 
and using (\ref{eq:(I-6)}), (\ref{eq:(IV-1)}) 
and 
\begin{equation} 
- \sqrt{2} C^{++}{}^{\alpha\beta} 
	( \sigma_\mu \bar{\sigma}^\nu \varepsilon )_{\alpha\beta} 
	(\xi^{(n)}_i \sigma_\nu \bar{\psi}_j ) \bar{\phi} 
= 
	(\xi^{(n+1)}_i \sigma_\mu \bar{\psi}_j ) (u^{+1})^2 
	, 
\end{equation} 
we can see that all the $O(C^2)$ terms are gauged away 
and the contributions to $\delta_\xi^* A_\mu$ come only 
from the first term and the order $C$ part of the second term 
in (\ref{eq:EqsToDetermineDeformedSUSY(1,1)}).

Eq.(\ref{eq:EqsToDetermineDeformedSUSY(0,2)}) 
have the same structure as eq.(\ref{eq:EqsToDetermineDeformedSUSY(1,1)}) 
in terms of the harmonic variables. 
Therefore, it is solved in a way similar 
to (\ref{eq:EqsToDetermineDeformedSUSY(1,1)}): 
The $O(C^2)$ terms in the r.h.s. of 
eq.(\ref{eq:EqsToDetermineDeformedSUSY(0,2)}) are completely gauged away, 
so that we need to consider only the first and the second term to determine 
$\delta_\xi^* \phi$.

In eq.(\ref{eq:EqsToDetermineDeformedSUSY(2,1)}), 
$\lambda^{(2,1)}_{\dot{\alpha}}$ is given by 
\begin{eqnarray} 
\lambda^{(2,1)}_{\dot{\alpha}} 
&=& 
	\sqrt{2} \partial_\mu 
		\left\{ 
		(\xi_i^{(1)} \sigma^\mu )_{\dot{\alpha}} 
		\bar{\phi} 
		\right\}  
		u^{-(i}u^{-1}u^{-1)}
		\nonumber\\ 
&&{} 
	+ \sum_{n=2}^{\infty} 
		2 \sqrt{2} \partial_\mu 
		\left\{ 
		(\xi_i^{(n)} \sigma^\mu )_{\dot{\alpha}} 
		\bar{\phi} 
		\right\}  
		\Biggl[
		\alpha^{(2,1)}_n 
			\overbrace{u^{+(1}\cdots u^{+1}}^{n-1} 
			\overbrace{u^{-i}u^{-1}\cdots u^{-1)}}^{n+2} 
		\nonumber\\ 
&& {} \qquad 
		+ \beta^{(2,1)}_n \epsilon^{i1} (u^{+1})^{n-2} (u^{-1})^{n+1}
		\ \Biggr] 
, 
\label{eq:SUSYLambda(2,1)}
\end{eqnarray} 
where 
\begin{equation} 
\alpha^{(2,1)}_n 
= 
	{1\over n+2} ( \alpha^{(1,0)}_n + \alpha^{(1,0)}_{n-1} ) 
	, 
	\quad 
\beta^{(2,1)}_n
= 
	{1\over n+1} 
		\left( 
		{1\over (2n+1)(2n-1)} \alpha^{(1,0)}_{n-1} 
		+ \beta^{(1,0)}_n 
		+ \beta^{(1,0)}_{n-1} 
		\right) 
	. 
\label{eq:NumSeries(2,1)}
\end{equation} 
We can easily find that 
in the r.h.s. of (\ref{eq:EqsToDetermineDeformedSUSY(2,1)}) 
the terms proportional to $(C_{11})^n$ ($n\ge 2$) 
are completely cancelled by this gauge parameter, 
so that there are no $O((C_{11})^2)$ terms in $\delta^*_\xi \bar{\psi}^i$.

Substituting (\ref{eq:SUSYLambda(0,1)}), (\ref{eq:SUSYLambda(1,0)}), 
(\ref{eq:SUSYLambda(1,1)}) and (\ref{eq:SUSYLambda(2,1)}) to 
eq.(\ref{eq:EqsToDetermineDeformedSUSY(1,2)}), 
we can check that the following form of the gauge parameter is 
sufficient 
: 
\begin{eqnarray} 
\lambda^{(1,2)}_\alpha  
&=& 
	\sum_{n=0}^{\infty} \chi_{\alpha ijk}{}_{(n)} 
		\overbrace{
		u^{+}{}^{(1} \dots u^{+}{}^{1} 
		}^{n}
		\overbrace{
		u^{-i} u^{-j} u^{-k} u^{-}{}^{1} \dots u^{-}{}^{1)}
		}^{n+3}
		\nonumber\\
&&{}
	+ \sum_{n=1}^{\infty} \chi_{\alpha ij}{}_{(n)}
		\overbrace{
		u^{+}{}^{(1} \dots u^{+}{}^{1} 
		}^{n-1}
		\overbrace{
		u^{-i} u^{-j} u^{-}{}^{1} \dots u^{-}{}^{1)}
		}^{n+2}
		\nonumber\\
&&{}
	+ \sum_{n=1}^{\infty} \chi_{\alpha i}{}_{(n)}
		\overbrace{
		u^{+}{}^{(1} \dots u^{+}{}^{1} 
		}^{n-1}
		\overbrace{
		u^{-i} u^{-}{}^{1} \dots u^{-}{}^{1)}
		}^{n+2}
	+ \sum_{n=2}^{\infty} \tilde{\chi}_{\alpha i}{}_{(n)}
		\overbrace{
		u^{+}{}^{(1} \dots u^{+}{}^{1} 
		}^{n-2}
		\overbrace{
		u^{-i} u^{-}{}^{1} \dots u^{-}{}^{1)}
		}^{n+1}
		\nonumber\\
&&{}
	+ \sum_{n=2}^{\infty} \chi_{\alpha}{}_{(n)}
		\overbrace{
		u^{+}{}^{(1} \dots u^{+}{}^{1} 
		}^{n-2}
		\overbrace{
		u^{-}{}^{1} \dots u^{-}{}^{1)}
		}^{n+1}
	+ \sum_{n=3}^{\infty} \tilde{\chi}_{\alpha}{}_{(n)}
		\overbrace{
		u^{+}{}^{(1} \dots u^{+}{}^{1} 
		}^{n-3}
		\overbrace{
		u^{-}{}^{1} \dots u^{-}{}^{1)}
		}^{n}
, 
\label{eq:SUSYLambda(1,2)}
\end{eqnarray} 
where the subscript $n$ denotes the power of $C_{11}$-dependence 
of each quantity.  
To see this is sufficient, we need (\ref{eq:(I-6)})%
--%
(\ref{eq:(II-5)}). 
Note that in the r.h.s. of eq.(\ref{eq:EqsToDetermineDeformedSUSY(1,2)}) 
the terms proportional to $(C_{11})^n$ ($n\ge 3$) are completely gauged away, 
so that there are no $O((C_{11})^3)$ terms in $\delta^*_\xi \psi^i$%
. 
In fact, in order to determine $\delta^*_\xi \psi^i$, only 
the precise forms of $\chi^\alpha_{ ijk}{}_{(0)}$ and $\chi^\alpha_{ ij}{}_{(1)}$ 
are needed%
: 
\begin{eqnarray} 
&& 
\chi^\alpha_{ ijk}{}_{(0)} 
= 
	2 \xi^\alpha_{ (i} D_{jk)} 
	, 
	\\
&& 
\chi^\alpha_{ ij}{}_{(1)} 
= 
	{32\over 15} 
		\left[ 
		\left\{ i (\bar{\psi}_i \bar{\psi}_j) 
		- \sqrt{2} D_{ij} \bar{\phi} 
		\right\}  
			(\xi_k \varepsilon C_{11})^\alpha 
		+ \Bigl( \mbox{cyclic permut. of $(ijk)$} \Bigr)
		\right] 
		\epsilon^{k1}
	. 
\end{eqnarray} 

Substituting (\ref{eq:SUSYLambda(1,0)}) 
and (\ref{eq:SUSYLambda(1,1)}) to 
eq.(\ref{eq:EqsToDetermineDeformedSUSY(2,2)}) 
and collecting $u^+$-independent terms, 
we obtain $\delta^*_{\xi} D^{ij}$. 
We do not need the precise form of $\lambda^{(2,2)}$.  
To determine $\delta^*_\xi D^{ij}$, 
we can use (\ref{eq:(II-5)}) and (\ref{eq:(IV-1)}).

\subsection{${\cal N}=(0,1/2)$ supersymmetry} 
\label{app:DeterminationOfN=(0,1/2)}

The equations to determine the deformed ${\cal N}=(0,1/2)$  
supersymmetry transformation are 
eqs.(\ref{eq:EqsToDetermineDeformedSUSY2(0,1)})--%
(\ref{eq:EqsToDetermineDeformedSUSY2(2,2)}) 
which are obtained from eq.(\ref{eq:DeformedSUSY(0,1/2)}) 
(note that here the expression (\ref{eq:AnalyticGaugeParam}) 
is used for the analytic gauge parameter $\Lambda'$): 
\begin{eqnarray}
0 
&=& 
	2 \sqrt{2} i \bar{\xi}^{+\dot{\alpha}} \phi 
	- \partial^{++} \lambda^{(0,1)}{}^{\dot{\alpha}} 
	- 2 \lambda^{(1,0)}{}^{\alpha} 
			(\varepsilon C^{++} \sigma^\mu )_{\alpha{\dot{\beta}}}
			\varepsilon^{{\dot{\beta}}{\dot{\alpha}}} 
			A_\mu 
	, 
	\label{eq:EqsToDetermineDeformedSUSY2(0,1)}
	\\
0 
&=& 
	- 2 i (\sigma^\mu \bar{\xi}^+)_\alpha A_\mu 
	- \partial^{++} \lambda^{(1,0)}_\alpha 
	- 2 \sqrt{2} (\varepsilon C^{++} \lambda^{(1,0)} )_\alpha \bar{\phi} 
	, 
	\label{eq:EqsToDetermineDeformedSUSY2(1,0)}
\end{eqnarray} 
\begin{eqnarray} 
\sqrt{2} i 
\delta^*_{\bar{\xi}} \phi 
&=& 
	4 i \lambda^{(1,0)}{}^{\alpha} (\varepsilon C^{++} \psi^i )_\alpha u_i^{-} 
	\nonumber\\ 
&&{} 
	+ {\sqrt{2} \over 3} \partial_\mu \lambda^{(2,0)} \partial_\nu \bar{\phi} \ 
			C^{\alpha_1 \beta_1}_{i_1 j_1} 
			C^{\alpha_2 \beta_2}_{i_2 j_2} 
			C^{\alpha_3 \beta_3}_{i_3 j_3} 
			\nonumber\\ 
&&{} 
			( 
			- u^{-i_1} u^{+i_2} u^{+i_3} \varepsilon_{\alpha_3 \alpha_2} 
				\sigma^\mu_{\alpha_1 \dot{\alpha}}
			+ u^{+i_1} u^{-i_2} u^{+i_3} \varepsilon_{\alpha_3 \alpha_1} 
				\sigma^\mu_{\alpha_2 \dot{\alpha}}
			- u^{+i_1} u^{+i_2} u^{-i_3} \varepsilon_{\alpha_1 \alpha_2} 
				\sigma^\mu_{\alpha_3 \dot{\alpha}}
			) 
			\nonumber\\ 
&&{} 
			( 
			- u^{-j_1} u^{+j_2} u^{+j_3} \varepsilon_{\beta_3 \beta_2} 
				\sigma^\nu_{\beta_1 \dot{\beta}}
			+ u^{+j_1} u^{-j_2} u^{+j_3} \varepsilon_{\beta_3 \beta_1} 
				\sigma^\nu_{\beta_2 \dot{\beta}}
			- u^{+j_1} u^{+j_2} u^{-j_3} \varepsilon_{\beta_1 \beta_2} 
				\sigma^\nu_{\beta_3 \dot{\beta}}
			) 
	\nonumber\\
&&{} 
	- \partial^{++} \lambda^{(0,2)} 
	- C^{++}{}^{\alpha\beta} 
			(\sigma^\nu \bar{\sigma}^\mu \varepsilon )_{\alpha\beta} 
			\lambda_\mu^{(1,1)} A_\nu 
	, 
	\label{eq:EqsToDetermineDeformedSUSY2(0,2)}
\end{eqnarray} 
\begin{eqnarray} 
- \sqrt{2} i 
\delta^*_{\bar{\xi}} \bar{\phi}
&=& 
	- 4 \bar{\xi}^+ \bar{\psi}^i u^{-}_i  
	- \partial^{++} \lambda^{(2,0)} 
	, 
	\label{eq:EqsToDetermineDeformedSUSY2(2,0)}
	\\
- 2 i 
\delta^*_{\bar{\xi}} A_\mu 
&=& 
	- 4 \psi^i \sigma_\mu \bar{\xi}^+ u^{-}_i 
	+ 4 i \lambda^{(1,0)}{}^{\alpha} 
			( \varepsilon C^{++} \sigma_{\mu} 
			\bar{\psi}^i )_\alpha u_i^{-} 
	- 2 C^{++}{}^{\alpha\beta} 
			(\sigma^{\nu} 
			\bar{\sigma}_{\mu} \varepsilon )_{\alpha\beta}  
			A_{\nu} \lambda^{(2,0)} 
	\nonumber\\
&&{} 
	- \partial^{++} \lambda_\mu^{(1,1)} 
	- \sqrt{2} C^{++}{}^{\alpha\beta} 
			( \sigma_\mu \bar{\sigma}^\nu \varepsilon)_{\alpha\beta}
			\lambda_\nu^{(1,1)} \bar{\phi}
	, 
	\label{eq:EqsToDetermineDeformedSUSY2(1,1)}
	\\
4 
\delta^*_{\bar{\xi}} \psi_{\alpha}^i u_i^{-} 
&=& 
	2 \sqrt{2} (\sigma^\mu \bar{\xi}^-)_\alpha \partial_\mu \phi 
			\nonumber\\
&&{} 
	- i (\sigma^\nu \partial_\nu \lambda^{(0,1)})_\alpha 
	- 2 \sqrt{2} i ( \varepsilon C^{+-} 
			\sigma^\nu \partial_\nu \lambda^{(0,1)} )_\alpha \bar{\phi} 
			\nonumber\\
&&{} 
	- 6 i (\varepsilon C^{++} \lambda^{(1,0)})_{\alpha} D^{ij} u_i^{-} u_j^{-} 
	+ 2 i \lambda^{(1,0)}{}^{\beta} 
			(\varepsilon C^{+-} \sigma^\nu 
			\bar{\sigma}^\mu \varepsilon)_{\beta\alpha} 
			\partial_\nu A_\mu 
			\nonumber\\
&&{} 
	+ 2 i \partial_\nu \lambda^{(1,0)}_{\alpha} C^{+-}{}^{\beta\gamma} 
			(\sigma^\mu \bar{\sigma}^\nu \varepsilon)_{\beta\gamma} 
			A_\mu 
	+ 8 i (\varepsilon C^{++} \psi^{-})_{\alpha} \lambda^{(2,0)} 
			\nonumber\\
&&{} 
	- 2 i (\sigma^\mu \bar{\psi}^i )_{\alpha} u_i^{-} 
			C^{++}{}^{\gamma\delta} 
			(\sigma_\mu \bar{\sigma}^\nu \varepsilon)_{\gamma\delta}
			\lambda_\nu^{(1,1)} 
			\nonumber\\
&&{} 
	- \partial^{++} \lambda^{(1,2)}_\alpha 
	- 2 \sqrt{2} ( \varepsilon C^{++} \lambda^{(1,2)} )_\alpha \bar{\phi} 
	- 2 
		(\varepsilon C^{++} \sigma^\mu \lambda^{(2,1)} )_\alpha A_\mu 
	, 
	\label{eq:EqsToDetermineDeformedSUSY2(1,2)}
	\\
-4 
\delta^*_{\bar{\xi}} \bar{\psi}^{\dot{\alpha} i} u_i^{-}
&=& 
	6 \bar{\xi}^{+\dot{\alpha}} D^{ij}u^{-}_i u^{-}_j
	- 2 (\bar{\sigma}^\nu \sigma^\mu \bar{\xi}^- )^{\dot{\alpha}} 
			\partial_\mu A_\nu 
	+ i \partial_\mu \lambda^{(1,0)}{}^\alpha 
			\sigma^\mu_{\alpha{\dot{\beta}}} 
			\varepsilon^{\dot{\beta}\dot{\alpha}} 
			\nonumber\\
&&{} 
	+ 2 \sqrt{2} i \partial_\nu 
		\left\{ 
		\lambda^{(1,0)}{}^{\alpha} 
			(\varepsilon C^{+-} \sigma^\nu )_{\alpha{\dot{\beta}}} 
			\varepsilon^{{\dot{\beta}}{\dot{\alpha}}} 
			\bar{\phi} 
		\right\} 
	- \partial^{++} \lambda^{(2,1)}{}^{\dot{\alpha}} 
	, 
	\label{eq:EqsToDetermineDeformedSUSY2(2,1)}
	\\ 
3
\delta^*_{\bar{\xi}} D^{ij} u_i^{-} u_j^{-} 
&=& 
	- 4 i \bar{\xi}^- \bar{\sigma}^\mu \partial_\mu \psi^i u^{-}_i 
	+ 4 \partial_\nu 
		\left\{ 
		\lambda^{(1,0)}{}^{\alpha} 
			(\varepsilon C^{+-} \sigma^\nu \bar{\psi}^i )_{\alpha} u_i^{-} 
		\right\} 
			\nonumber\\
&&{} 
	+ 2 i C^{+-}{}^{\alpha\beta} 
		(\sigma^{\mu} 
		\bar{\sigma}^{\nu} \varepsilon 
		)_{\alpha\beta} 
		\partial_\nu (\lambda^{(2,0)} A_\mu) 
			\nonumber\\
&&{} 
	- i \partial^\mu \lambda_\mu^{(1,1)} 
	- \sqrt{2} i C^{+-}{}^{\alpha\beta} 
			( \sigma^\mu \bar{\sigma}^\nu \varepsilon )_{\alpha\beta} 
			\partial_\nu ( \lambda_\mu^{(1,1)} \bar{\phi} )
	- \partial^{++} \lambda^{(2,2)}
. 
\label{eq:EqsToDetermineDeformedSUSY2(2,2)}
\end{eqnarray} 
Here we should understand that $\bar{\xi}^1$ is implicitly set to be zero. 
Note that under the restriction (\ref{eq:reduced1}), we have 
$C^{+-}{}^{\alpha\beta} = C_{11}^{\alpha\beta} u^{+1}u^{-1}$ and so on.

Eq.(\ref{eq:EqsToDetermineDeformedSUSY2(1,0)}) is solved by 
\begin{eqnarray} 
\lambda^{(1,0)}{}^\alpha 
&=& 
	- 2 i (\bar{\xi}_i \bar{\sigma}^\mu )^\alpha A_\mu \ u^{-i} 
	+ \sum_{n=1}^{\infty} 
		(-2i) (\bar{\xi}_i \bar{\sigma}^\mu \varepsilon {\cal C}^{(n)} )^\alpha 
		A_\mu (2\sqrt{2} \bar{\phi})^n 
		\nonumber\\ 
&&{} \qquad \times 
		\biggl[
		\alpha^{(1,0)}_n 
			\overbrace{u^{+(1}\cdots u^{+1}}^{n} 
			\overbrace{u^{-i}u^{-1}\cdots u^{-1)}}^{n+1} 
		+ \beta^{(1,0)}_n \epsilon^{i1} (u^{+1})^{n-1} (u^{-1})^n
		\biggr]
, 
\label{eq:SUSY2Lambda(1,0)}
\end{eqnarray} 
where 
\begin{equation} 
{\cal C}^{(n)}{}^{\alpha\beta}  
\equiv 
	( 
	\overbrace{ C_{11} \varepsilon C_{11} \cdots \varepsilon C_{11} }^{n}
	)^{\alpha\beta} 
\end{equation} 
and $\alpha^{(1,0)}_n, \beta^{(1,0)}_n$ are given in (\ref{eq:NumSeries(1,0)}). 
To check this, 
we need (\ref{eq:(I-6)}) and (\ref{eq:(IV-1)}).

Then from eq.(\ref{eq:EqsToDetermineDeformedSUSY2(0,1)}) we find 
\begin{eqnarray} 
\lambda^{(0,1)}_{\dot{\alpha}} 
&=& 
	- 2 \sqrt{2} i \bar{\xi}_{\dot{\alpha}}{}_i {\phi} \ u^{-i} 
	+ \sum_{n=1}^{\infty} 
		(- 4i) (\bar{\xi}_i \bar{\sigma}^\mu \varepsilon {\cal C}^{(n)} 
		\sigma^\nu )_{\dot{\alpha}} A_\mu A_\nu (2\sqrt{2}\bar{\phi})^{n-1} 
		\nonumber\\ 
&&{} \qquad \qquad \times 
		\biggl[ 
		\alpha^{(1,0)}_{n} 
				\overbrace{u^{+(1}\cdots u^{+1}}^{n} 
				\overbrace{u^{-i}u^{-1}\cdots u^{-1)}}^{n+1} 
		+ \beta^{(1,0)}_n \epsilon^{i1} 
			(u^{+1})^{n-1} (u^{-1})^n
		\biggr] 
	. 
	\nonumber\\ 
&& 
\label{eq:SUSY2Lambda(0,1)}
\end{eqnarray} 

From eq.(\ref{eq:EqsToDetermineDeformedSUSY2(2,0)}), 
we find 
$\delta_{\bar{\xi}}^* \bar{\phi} = \sqrt{2} i \bar{\xi}_i \bar{\psi}^i$ and 
\begin{equation} 
\lambda^{(2,0)} 
= 
	- 2 \bar{\xi}_k \bar{\psi}_l u^{-(k} u^{-l)} 
	. 
\label{eq:SUSY2Lambda(2,0)}
\end{equation}

In eq.(\ref{eq:EqsToDetermineDeformedSUSY2(1,1)}), 
$\lambda^{(1,1)}_\mu$ is given by 
\begin{eqnarray} 
\lambda^{(1,1)}_\mu 
&=& 
	-2 \psi_i \sigma_\mu \bar{\xi}_j \ u^{-(i}u^{-j)} 
		\nonumber\\ 
&&{} 
	+ \left( 
		\lambda^{(1,1)}_{\mu\ ij}{}_{(1)} 
		+ \tilde{\lambda}^{(1,1)}_{\mu\ ij}{}_{(1)} 
	\right) 
		\left[ 
		2 u^{+(1}u^{-i}u^{-j}u^{-1)}
		+ \frac32  
			\Bigl\{ 
			\epsilon ^{i1} 
				u^{-(j}u^{-1)} 
			+ ( i \leftrightarrow j ) 
			\ \Bigr\}
		\right] 
		\nonumber\\ 
&&{} 
	+ \sum_{n=2}^{\infty} 
		\lambda^{(1,1)}_{\mu\ ij}{}_{(n)} 
		\biggl[ 
		\alpha^{(1,1)}_n 
			\overbrace{u^{+(1}\cdots u^{+1}}^{n} 
			\overbrace{u^{-i}u^{-j}u^{-1}\cdots u^{-1)}}^{n+2} 
		\nonumber\\ 
&&{} \qquad 
		+ \beta^{(1,1)}_n 
			\frac12 
			\Bigl\{ 
			\epsilon ^{i1} 
				\overbrace{u^{+(1}\cdots u^{+1}}^{n-1} 
				\overbrace{u^{-j}u^{-1}\cdots u^{-1)}}^{n+1} 
			+ ( i \leftrightarrow j ) 
			\ \Bigr\}
		\nonumber\\ 
&&{} \qquad 
		+ \gamma^{(1,1)}_n 
			\epsilon ^{i1} \epsilon ^{j1} 
			(u^{+1})^{n-2} (u^{-1})^n
		+ \delta^{(1,1)}_n 
			\epsilon^{ji} 
			(u^{+1})^{n-1} (u^{-1})^{n+1} 
		\biggr] 
		\nonumber\\ 
&&{} 
	+ \sum_{n=2}^{\infty} 
		\tilde{\lambda}^{(1,1)}_{\mu\ ij}{}_{(n)} 
		\biggl[ 
		\tilde{\alpha}^{(1,1)}_n 
			\overbrace{u^{+(1}\cdots u^{+1}}^{n} 
			\overbrace{u^{-i}u^{-j}u^{-1}\cdots u^{-1)}}^{n+2} 
		\nonumber\\ 
&&{} \qquad 
		+ \tilde{\beta}^{(1,1)}_n 
			\frac12 
			\Bigl\{ 
			\epsilon ^{i1} 
				\overbrace{u^{+(1}\cdots u^{+1}}^{n-1} 
				\overbrace{u^{-j}u^{-1}\cdots u^{-1)}}^{n+1} 
			+ ( i \leftrightarrow j ) 
			\ \Bigr\}
		\nonumber\\ 
&&{} \qquad 
		+ \tilde{\gamma}^{(1,1)}_n 
			\epsilon ^{i1} \epsilon ^{j1} 
			(u^{+1})^{n-2} (u^{-1})^n
		\biggr] 
, 
\label{eq:SUSY2Lambda(1,1)} 
\end{eqnarray} 
where 
\begin{eqnarray} 
&& 
\lambda^{(1,1)}_{\mu\ ij}{}_{(n)} 
\equiv 
	- \frac43 (\bar{\xi}_{(i} \bar{\sigma}^\nu \varepsilon {\cal C}^{(n)} 
		\sigma_\mu \bar{\psi}_{j)} ) 
		A_\nu ( 2\sqrt{2} \bar{\phi} )^{n-1} 
	, \nonumber \\ 
&& 
\tilde{\lambda}^{(1,1)}_{\mu\ ij}{}_{(n)} 
\equiv 
	\frac23 {\cal C}^{(n)}{}^{\alpha\beta} 
		(\sigma^\nu \bar{\sigma}_\mu \varepsilon)_{\alpha\beta}
		A_\nu ( 2\sqrt{2} \bar{\phi} )^{n-1} \bar{\xi}_{(i}\bar{\psi}_{j)} 
	- \frac13 
		( \psi_{(i} \varepsilon {\cal C}^{(n)} \sigma_\mu \bar{\xi}_{j)} ) 
		( 2\sqrt{2} \bar{\phi} )^{n} 
	, 
\end{eqnarray} 
and 
\begin{eqnarray} 
&& 
(\alpha^{(1,1)}_2, \beta^{(1,1)}_2, \gamma^{(1,1)}_2, \delta^{(1,1)}_2)
= \left(\matrix{ \frac54 , & \frac {10}3, 
	& 2, & -\frac12 }\right)
	, \quad 
\alpha^{(1,1)}_{n} 
= 
	{1\over n+1} \left( 6 \alpha^{(1,0)}_{n-1} + \alpha^{(1,1)}_{n-1} \right) 
	, \nonumber\\ 
&& 
\beta^{(1,1)}_{n} 
= 
	{1\over n+1} 
		\left( 
		{6n\over 2n-1} \alpha^{(1,0)}_{n-1} 
		+ 6 \beta^{(1,0)}_{n-1} 
		+ {1\over n} \alpha^{(1,1)}_{n-1} 
		+ \beta^{(1,1)}_{n-1} 
		\right) 
	, \nonumber\\ 
&& 
\gamma^{(1,1)}_{n} 
= 
	{1\over n} 
		\left( 
		{6n\over (2n+1)(2n-1)} \alpha^{(1,0)}_{n-1} 
		+ 3 \beta^{(1,0)}_{n-1} 
		\right. \nonumber\\ 
&&\left. {} \qquad 
		+ {n+1 \over (2n+1)2n(2n-1)} \alpha^{(1,1)}_{n-1} 
		+ {1\over 2n-2} \beta^{(1,1)}_{n-1} 
		+ \gamma^{(1,1)}_{n-1} 
		\right) 
	, \nonumber\\ 
&& 
\delta^{(1,1)}_{n} 
= 
	{1\over n+1} 
		\left( 
		{3(n-1)\over 2n-1} \alpha^{(1,0)}_{n-1} 
		- 3 \beta^{(1,0)}_{n-1} 
		+ \delta^{(1,1)}_{n-1} 
		\right) 
	, \\ 
&& 
(\tilde{\alpha}^{(1,1)}_2, \tilde{\beta}^{(1,1)}_2, 
	\tilde{\gamma}^{(1,1)}_2)
= \left(\matrix{ \frac12 , & \frac43, & \frac45}\right)
	, \nonumber\\ 
&& 
\tilde{\alpha}^{(1,1)}_{n} 
= 
	{1\over n+2} \tilde{\alpha}^{(1,1)}_{n-1} 
	, \quad 
\tilde{\beta}^{(1,1)}_{n} 
= 
	{1\over n+1} 
		\left( 
		{1\over n} \tilde{\alpha}^{(1,1)}_{n-1} 
		+ \tilde{\beta}^{(1,1)}_{n-1} 
		\right) 
	, \nonumber\\ 
&& 
\tilde{\gamma}^{(1,1)}_{n} 
= 
	{1\over n} 
		\left( 
	{n+1 \over (2n+1)2n(2n-1)} \tilde{\alpha}^{(1,1)}_{n-1} 
		+ {1\over 2n-2} \tilde{\beta}^{(1,1)}_{n-1} 
		+ \tilde{\gamma}^{(1,1)}_{n-1} 
		\right) 
	. 
\end{eqnarray} 
Substituting eq.(\ref{eq:SUSY2Lambda(1,1)}) to the r.h.s. of 
eq.(\ref{eq:EqsToDetermineDeformedSUSY2(1,1)}) 
and using (\ref{eq:(I-6)}), (\ref{eq:(IV-1)}) 
and 
\begin{equation} 
- \sqrt{2} C_{11}^{\alpha\beta}  
	( \sigma_\mu \bar{\sigma}^\nu \varepsilon )_{\alpha\beta} 
	\tilde{\lambda}^{(1,1)}_{\nu\ ij}{}_{(n)} \bar{\phi} 
= 
	\tilde{\lambda}^{(1,1)}_{\mu\ ij}{}_{(n+1)} 
	, 
\end{equation} 
we can see that all the $O((C_{11})^2)$ terms are gauged away 
and the contributions to $\delta_{\bar{\xi}}^* A_\mu$ come only 
from the $(C_{11})^1$ part of the second term besides the first term.

The $C^3$ term in (\ref{eq:EqsToDetermineDeformedSUSY2(0,2)}) becomes 
$$
	+ {16\sqrt{2} \over 3} \det C_{11} C_{11}^{\mu\nu} 
		\partial_\mu \lambda^{(2,0)} \partial_\nu \bar{\phi} \ 
		(u^{+1})^4 (u^{-1})^2 
, 
$$
which will be gauged away (because of the excessive number of $u^{+}$). 
As a result, we can regard eq.(\ref{eq:EqsToDetermineDeformedSUSY2(0,2)}) 
as having the same structure 
as eq.(\ref{eq:EqsToDetermineDeformedSUSY2(1,1)}) 
in terms of the harmonic variables. 
Therefore, it is solved in a way similar 
to (\ref{eq:EqsToDetermineDeformedSUSY2(1,1)}): 
The $O((C_{11})^2)$ terms in the r.h.s. of 
eq.(\ref{eq:EqsToDetermineDeformedSUSY2(0,2)}) are completely gauged away, 
and the relevant contributions to 
$\delta_{\bar{\xi}}^* \phi$ are coming from each $O((C_{11})^0)$ part of 
$\lambda^{(1,0)}_\alpha$ and $\lambda^{(1,1)}_\mu$.

In eq.(\ref{eq:EqsToDetermineDeformedSUSY2(2,1)}), 
$\lambda^{(2,1)}_{\dot{\alpha}}$ is given by 
\begin{eqnarray} 
\lambda^{(2,1)}_{\dot{\alpha}} 
&=& 
	-2 \bar{\xi}_{\dot{\alpha}}{}_{(i} D_{jk)}  
		u^{-(i}u^{-j}u^{-k)}
	- (\bar{\xi}_i \bar{\sigma}^\nu \varepsilon 
		C_{11} \sigma^\mu )_{\dot{\alpha}} \partial_\mu ( A_\nu \bar{\phi} ) 
		u^{-(i}u^{-1}u^{-1)}
		\nonumber\\ 
&&{} 
	- \sum_{n=2}^{\infty} 
		2 (\bar{\xi}_i \bar{\sigma}^\nu \varepsilon 
		{\cal C}^{(n)} \sigma^\mu )_{\dot{\alpha}} 
		\partial_\mu \left\{ A_\nu ( 2\sqrt{2} \bar{\phi} )^{n} \right\} 
		\Biggl[
		\alpha^{(2,1)}_n 
			\overbrace{u^{+(1}\cdots u^{+1}}^{n-1} 
			\overbrace{u^{-i}u^{-1}\cdots u^{-1)}}^{n+2} 
		\nonumber\\ 
&& {} \qquad 
		+ \beta^{(2,1)}_n \epsilon^{i1} (u^{+1})^{n-2} (u^{-1})^{n+1}
		\ \Biggr] 
, 
\label{eq:SUSY2Lambda(2,1)}
\end{eqnarray} 
where $\alpha^{(2,1)}_n, \beta^{(2,1)}_n$ are given in (\ref{eq:NumSeries(2,1)}). 
We can easily find that 
in the r.h.s. of (\ref{eq:EqsToDetermineDeformedSUSY2(2,1)}) 
the terms proportional to $(C_{11})^n$ ($n\ge 2$) 
are completely cancelled by this gauge parameter, 
so that there are no $O((C_{11})^2)$ terms in $\delta^*_\xi \bar{\psi}^i$.

Substituting (\ref{eq:SUSY2Lambda(0,1)}), (\ref{eq:SUSY2Lambda(1,0)}), 
(\ref{eq:SUSY2Lambda(2,0)}), (\ref{eq:SUSY2Lambda(1,1)}) 
and (\ref{eq:SUSY2Lambda(2,1)}) into 
eq.(\ref{eq:EqsToDetermineDeformedSUSY2(1,2)}), 
we can check that the gauge parameter $\lambda^{(1,2)}_\alpha$ 
having the same form as (\ref{eq:SUSYLambda(1,2)}) is sufficient. 
To see it is sufficient, we need (\ref{eq:(I-6)})%
--%
(\ref{eq:(II-5)}). 
In fact, in order to determine $\delta^*_{\bar{\xi}} \psi^i$, only 
the precise forms of $\chi_{\alpha ijk}{}_{(0)}$ and $\chi_{\alpha ij}{}_{(1)}$ 
are needed as in the ${\cal N}=(1,0)$ case%
: 
\begin{equation} 
\chi_{\alpha ijk}{}_{(0)} 
= 
	0 
	, 
	\quad 
\chi_{\alpha ij}{}_{(1)} 
= 
	{32\over 5} \left[ 
	D_{(ij} ( \bar{\xi}_{k)} \bar{\sigma}^\mu \varepsilon C_{11} )^\beta 
		\varepsilon_{\alpha\beta} A_\mu 
	+ 2 i ( \varepsilon C_{11} \psi_{(i} )_\alpha \bar{\xi}_j 
		\bar{\psi}_{k)} 
	\right] 
	\epsilon^{k1} 
	.  
\end{equation} 

Substituting (\ref{eq:SUSY2Lambda(1,0)}), (\ref{eq:SUSY2Lambda(2,0)}) 
and (\ref{eq:SUSY2Lambda(1,1)}) to 
eq.(\ref{eq:EqsToDetermineDeformedSUSY2(2,2)}) 
and collecting $u^+$-independent terms, 
we obtain $\delta^*_{\bar{\xi}} D^{ij}$. 
We do not need the precise form of $\lambda^{(2,2)}$. 
To determine $\delta^*_{\bar{\xi}} D^{ij}$, 
we can use (\ref{eq:(II-5)}) and (\ref{eq:(IV-1)}). 

The ${\cal N}=(0,1/2)$ supersymmetry transformation 
can be read from the following result 
by setting $\bar{\xi}^1=0$:  
\begin{eqnarray} 
\delta_{\bar{\xi}}^* \phi 
&=& 
	c_{1} i ( \bar{\xi}^1 \bar{\sigma}^\mu \varepsilon C_{11} \psi^1 ) 
		A_\mu 
	,
	\label{eq:N=(0,1):(0,2)}
	\\ 
\delta_{\bar{\xi}}^* \bar{\phi}
&=& 
	- \sqrt{2} i \bar{\xi}^i \bar{\psi}_i 
	,
	\label{eq:N=(0,1):(2,0)}
	\\ 
\delta_{\bar{\xi}}^* A_\mu 
&=& 
	i \bar{\xi}^i \bar{\sigma}_\mu \psi_i 
	+ c_{2} i (\bar{\xi}^1 \bar{\sigma}^\nu \varepsilon C_{11} \sigma_\mu 
		\bar{\psi}^1 ) A_\nu 
	\nonumber\\ 
&&{} 
	+ c_{3} i C_{11}^{\alpha\beta} 
		(\sigma^\nu \bar{\sigma}_\mu \varepsilon )_{\alpha\beta} 
		\bar{\xi}^1 \bar{\psi}^1 A_\nu 
	+ c_{4} i 
		( \bar{\xi}^1 \bar{\sigma}_\mu \varepsilon C_{11} \psi^1 ) 
		\bar{\phi} 
	, 
	\label{eq:N=(0,1):(1,1)}
	\\ 
\delta_{\bar{\xi}}^* \psi^{\alpha i} 
&=& 
	- \sqrt{2} ( \bar{\xi}^i \bar{\sigma}^\mu )^\alpha \partial_\mu \phi 
	- 2 D^{(i1} ( \bar{\xi}^{1)} \bar{\sigma}^\mu \varepsilon C_{11} )^\alpha 
		A_\mu 
	+ 4 i \bar{\xi}^{(i} \bar{\psi}^{1} ( \psi^{1)} \varepsilon C_{11} )^\alpha  
	\nonumber\\ 
&&{} 
	+ \delta^i_2 
		\left[ 
		c_{5} ( \bar{\xi}^1 \bar{\sigma}^\mu \varepsilon C_{11} )^\alpha 
			\partial_\mu \phi \bar{\phi} 
		+ ( \bar{\xi}^1 \bar{\sigma}^\mu \varepsilon C_{11} 
			\sigma^\nu \bar{\sigma}^\rho )^\alpha 
			\Bigl\{ 
			c_{6} \partial_\rho ( A_\mu A_\nu ) 
			+ c_{7} A_\mu \partial_\nu A_\rho 
			\Bigr\} 
			\right. \nonumber\\ 
&&\left. {} \qquad 
		+ c_{8} ( \bar{\xi}^1 \bar{\sigma}^\mu )^\alpha C_{11}^{\beta\gamma} 
			( \sigma^\nu \bar{\sigma}^\rho \varepsilon )_{\beta\gamma}  
			\partial_\rho A_\mu A_\nu 
		\right] 
	\nonumber\\ 
&&{} 
	+ 
	\delta^i_2 \det C_{11} 
		\left[ 
		( \bar{\xi}^1 \bar{\sigma}^\mu )^\alpha 
			\left\{ 
			c_{9} D^{11} A_\mu \bar{\phi} 
			+ c_{10} i \bar{\psi}^1 \bar{\psi}^1 A_\mu 
			\right\} 
		+ c_{11} i \bar{\xi}^1 \bar{\psi}^1 
			\psi^{\alpha 1} \bar{\phi} 
		\right] 
	, 
	\label{eq:N=(0,1):(1,2)}
	\\ 
\delta_{\bar{\xi}}^* \bar{\psi}_{\dot{\alpha} i} 
&=& 
	( \bar{\xi}_i \bar{\sigma}^{\mu\nu} )_{\dot{\alpha}} F_{\mu\nu} 
	- \bar{\xi}_{\dot{\alpha}}^{j} D_{ij} 
	+ \delta_i^1 c_{12} 
		\partial_\mu \left[ 
			( \bar{\xi}^1 \bar{\sigma}^\nu 
			\varepsilon C_{11} \sigma^\mu )_{\dot{\alpha}} 
			A_\nu \bar{\phi} 
		\right] 
	, 
	\label{eq:N=(0,1):(2,1)}
	\\ 
\delta_{\bar{\xi}}^* D^{11} 
&=& 
	c_{13} i \partial_\mu ( \bar{\xi}^1 \bar{\sigma}^\mu \psi^1 ) 
	, 
	\label{eq:N=(0,1):(2,2)}
	\nonumber\\ 
\delta_{\bar{\xi}}^* D^{12} 
&=& 
	- 2 i \partial_\mu ( \bar{\xi}^{(1} \bar{\sigma}^\mu \psi^{2)} ) 
	\nonumber\\ 
&&{} 
	+ i \partial_\mu 
		\left[
		c_{14} ( \bar{\xi}^1 \bar{\sigma}^\nu \varepsilon C_{11} \sigma^\mu 
			\bar{\psi}^1 ) A_\nu 
		+ c_{15} C_{11}^{\alpha\beta} 
			(\sigma^\nu \bar{\sigma}^\mu \varepsilon)_{\alpha\beta} 
			A_\nu \bar{\xi}^1 \bar{\psi}^1 
		+ c_{16} ( \bar{\xi}^1 \bar{\sigma}^\mu \varepsilon C_{11} 
			\psi^1 ) \bar{\phi} 
		\right] 
	,
	\nonumber \\ 
\delta_{\bar{\xi}}^* D^{22} 
&=& 
	- 2 i \partial_\mu ( \bar{\xi}^2 \bar{\sigma}^\mu \psi^2 ) 
	- \frac 43 i \partial_\mu 
		\left[
		2 ( \bar{\xi}^{(1} \bar{\sigma}^\nu \varepsilon C_{11} \sigma^\mu 
			\bar{\psi}^{2)} ) A_\nu 
			\right. \nonumber\\ 
&&\left. {} \qquad 
		- C_{11}^{\alpha\beta} 
			(\sigma^\nu \bar{\sigma}^\mu \varepsilon)_{\alpha\beta} 
			A_\nu \bar{\xi}^{(1} \bar{\psi}^{2)} 
		+ \sqrt{2} ( \bar{\xi}^{(1} \bar{\sigma}^\mu \varepsilon C_{11} 
			\psi^{2)} ) \bar{\phi} 
		\right] 
	\nonumber\\ 
&&{} 
	+ i \det C_{11} \partial_\mu 
		\left[
		c_{17} ( \bar{\xi}^{1} \bar{\sigma}^\nu \sigma^\mu 
			\bar{\psi}^{1} ) A_\nu \bar{\phi} 
		+ c_{18} \bar{\xi}^{1} \bar{\psi}^{1} A^\mu \bar{\phi} 
		+ c_{19} ( \bar{\xi}^{1} \bar{\sigma}^\mu \psi^{1} ) 
			\bar{\phi}^2 
		\right] 
, 
\end{eqnarray} 
where $c_i$'s are certain constants.

\end{document}